\documentclass[pra,showpacs,nobalancelastpage]{revtex4}
\usepackage{amssymb}
\usepackage{amsmath}
\usepackage{graphicx}

\begin{document}

\title{Quantum Teleportation with Continuous Variables: a survey }
\author{Stefano Pirandola and Stefano Mancini}

\affiliation{Dipartimento di Fisica, Universit\`a di Camerino, via
Madonna delle Carceri, I-62032 Camerino, Italy}
\date{\today}

\begin{abstract}
Very recently we have witnessed a new development of quantum
information, the so-called continuous variable (CV) quantum
information theory. Such a further development has been mainly due
to the experimental and theoretical advantages offered by CV
systems, i.e., quantum systems described by a set of observables,
like position and momentum, which have a continuous spectrum of
eigenvalues. According to this novel trend, quantum information
protocols like quantum teleportation have been suitably extended
to the CV framework. Here, we briefly review some mathematical
tools relative to CV systems and we consequently develop the
concepts of quantum entanglement and teleportation in the CV
framework, by analogy with the qubit-based approach. Some
connections between teleportation fidelity and entanglement
properties of the underlying quantum channel are inspected. Next,
we face the study of CV quantum teleportation networks where more
users share a multipartite state and an arbitrary pair of them
performs quantum teleportation. In this context, we show
alternative protocols and we investigate the optimal strategy that
maximizes the performance of the network.
\end{abstract}


\maketitle

\tableofcontents

\section{Introduction}

Ideally, quantum teleportation consists in the perfect transfer of an
unknown quantum state $\rho _{in}$ from a sender, usually called Alice, to a
remote receiver, usually called Bob \cite{TeleBennett}. To accomplish this
task, Alice divides the full information of $\rho _{in}$ into two parts, one
classical and the other non-classical, and send them to Bob through two
different channels: the classical channel and the quantum channel. The first
one can be a telegraph wire, while the latter one is a bipartite quantum
system with strong long-range correlations and which is distributed in
advance between the two parties. After these transmissions, Bob can act on
the quantum channel reconstructing a perfect replica of $\rho _{in}$, while
the original Alice's state is destroyed during the process (as a direct
consequence of the no-cloning theorem \cite{NoCloning}). The net result of
the teleportation is the removal of $\rho _{in}$ from Alice's side and its
appearance in Bob's side some time later, just the time needed by a
classical message to go from Alice to Bob.

The depicted general protocol for quantum teleportation is based on a
fundamental ingredient: the quantum channel which\ is shared by the two
parties. It consists in a bipartite state $\rho _{AB}$ of two distant
systems $A$ and $B$, possessed by Alice and Bob, respectively, and whose
correlations are exploited to teleport the input state. This is possible by
mixing the input state with the quantum channel at Alice's side (system $A$)
via a suitable collective measurement, called \emph{Bell measurement}, whose
projective effect causes the flow of quantum information to system $B$ via
the long-range correlations. The result of the measurement is then sent to
Bob through the classical channel, and Bob uses this additional classical
information to completely retrieve the input state from his system $B$. This
is possible by performing a unitary transformation which is conditioned to
the particular outcome of the Bell measurement.

\begin{figure}[tbph]
\vspace{-0.5cm}
\par
\begin{center}
\includegraphics[width=0.55\textwidth]{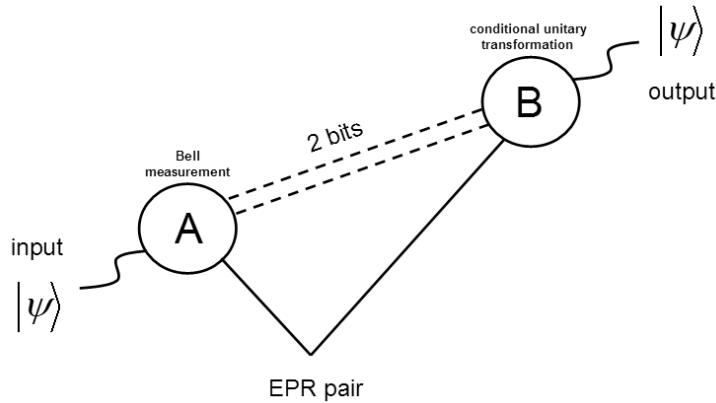}
\end{center}
\par
\vspace{-0.4cm} \caption{Teleportation protocol. Alice (A) and Bob
(B) share a quantum channel given by an EPR pair, and Alice wants
to teleport an unknown input state $\left| \protect\psi
\right\rangle $ to Bob. To accomplish this task, Alice performs a
Bell measurement on both her input and EPR qubits, and
communicates the outcome to Bob through a classical channel
(dashed lines). After this transmission, Bob applies a suitable
unitary transformation (conditioned to the measurement result) to
his EPR qubit and automatically reconstructs the input state
$\left| \protect\psi \right\rangle $ at the output.} \label{tele}
\end{figure}

Bennett \textit{et al. }\cite{TeleBennett} first show how the strong
correlations between a pair of qubits $A$ and $B$, prepared in the
Einstein-Podolsky-Rosen\emph{\ }(EPR)\ state
\begin{equation}
\left| \Psi _{AB}^{-}\right\rangle \equiv \left( \left| 0\right\rangle
_{A}\left| 1\right\rangle _{B}-\left| 1\right\rangle _{A}\left|
0\right\rangle _{B}\right) /\sqrt{2}  \label{EPR_state_qubits}
\end{equation}
can assist the perfect teleportation of an unknown pure quantum state $%
\left| \psi \right\rangle _{in}=a\left| 0\right\rangle _{in}+b\left|
1\right\rangle _{in}$ from Alice to Bob (see Fig.~\ref{tele}). The initial
total state $|\Psi _{\text{tot}}\rangle \equiv |\psi \rangle _{in}\otimes
|\Psi _{AB}^{-}\rangle $, involves no correlations between the unknown qubit
and the EPR\ pair. In order transmit the information through the quantum
channel, Alice couples the input qubit with the qubit $A$ of the EPR pair by
performing a Bell measurement. This a joint measurement over the qubits $in$
and $A$ given by the projection on the orthonormal basis
\begin{eqnarray}
\left| \Psi ^{\pm }\right\rangle &\equiv &\left( \left| 0\right\rangle
_{in}\left| 1\right\rangle _{A}\pm \left| 1\right\rangle _{in}\left|
0\right\rangle _{A}\right) /\sqrt{2}~,  \label{BellBasis2} \\
\left| \Phi ^{\pm }\right\rangle &\equiv &\left( \left| 0\right\rangle
_{in}\left| 0\right\rangle _{A}\pm \left| 1\right\rangle _{in}\left|
1\right\rangle _{A}\right) /\sqrt{2}~,
\end{eqnarray}
which is called \emph{Bell basis}. Since the total state $|\Psi _{\text{tot}%
}\rangle $ before the measurement can be expressed in terms of the Bell
basis as
\begin{equation}
\left| \Psi _{\text{tot}}\right\rangle =\frac{1}{2}\left[ \left| \Psi
^{-}\right\rangle (-\left| \psi \right\rangle _{B})+\left| \Psi
^{+}\right\rangle (-Z\left| \psi \right\rangle _{B})+\left| \Phi
^{-}\right\rangle (X\left| \psi \right\rangle _{B})+\left| \Phi
^{+}\right\rangle (XZ\left| \psi \right\rangle _{B})\right] ~,
\label{PsiBefore}
\end{equation}
where
\begin{equation*}
X\equiv \left(
\begin{array}{cc}
0 & 1 \\
1 & 0
\end{array}
\right) ,~Z\equiv \left(
\begin{array}{cc}
1 & 0 \\
0 & -1
\end{array}
\right) ,
\end{equation*}
are the Pauli operators, it follows that the four outcomes $\{\Psi ^{-},\Psi
^{+},\Phi ^{-},\Phi ^{+}\}$\ of the Bell measurement have all the
probabilities equal to $1/4$, and after the measurement, the qubit $B$ will
be projected into one of the following states
\begin{equation}
-\left| \psi \right\rangle _{B}~,~-Z\left| \psi \right\rangle _{B}~,~X\left|
\psi \right\rangle _{B}~,~XZ\left| \psi \right\rangle _{B}~.
\label{AfterBell}
\end{equation}
In every case, Bob's qubit $B$ will be described by a state simply related
to the original one $\left| \psi \right\rangle _{in}$. If now Alice
communicates to Bob the exact outcome of her measurement, Bob can
consequently apply the right unitary transformation on qubit $B$ which
retrieves the original input state. This is the classical part of the
protocol, i.e., the remaining information about the input state is
represented by two bits of classical information which are produced by the
measurement process. Exactly this amount of information has to be
communicated from Alice to Bob through a classical channel in order to
transmit the complete information about $\left| \psi \right\rangle _{in}$
and therefore complete the teleportation process.

Ideally, using a quantum channel as the previous EPR pair would allow for a
perfect quantum teleportation, where the output state is exactly equal to
the input one. However, in any real situation, Alice and Bob access limited
resources and the output state $\rho _{out}$ will be similar to input one $%
\rho _{in}$ by quantity called \emph{fidelity}, which is defined as
\begin{equation}
F\equiv \overline{\mathrm{Tr}(\sqrt{\rho _{in}}\rho _{out}\sqrt{\rho _{in}})}%
~,  \label{Fid_gen}
\end{equation}
where the bar denotes the average over many instances of the process (or the
outcomes of the Bell measurement). In particular, for a pure input state $%
\rho _{in}=|\psi \rangle _{in}\langle \psi |$, formula (\ref{Fid_gen})
simply becomes
\begin{equation}
F=\overline{\left\langle \psi \right| \rho _{out}\left| \psi \right\rangle }%
~.  \label{Fid_pure}
\end{equation}

One can ask if teleportation could be implemented classically, i.e., without
any shared quantum resource. For instance, in a purely classical strategy,
Alice can measure the input state on a fixed orthonormal basis and
communicate the outcomes to Bob, who, in turn, tries to reconstruct the
state from this classical information. However, Ref.~\cite{Popescu} proved
that the purely classical channel can give at most $F=2/3$, and therefore
the region $2/3<F\leq 1$ corresponds to a purely quantum teleportation.

In order to better understand this point, one has to introduce the concept
of entanglement between two quantum systems. Basically, a bipartite quantum
state $\rho _{AB}$ is entangled when it cannot be reproduced via local
operations and classical communication from the uncorrelated state $\rho
_{A}\otimes \rho _{B}$ \cite{Werner}. Entanglement is a physical resource
which can be quantified by introducing suitable entanglement measures \cite
{HOROmono}, and quantum teleportation is a process which exploits this
resource. When a bipartite system has a high degree of entanglement, the
correlations between its two subsystems $A$ and $B$ are so strong that they
are not reproducible by any classical means. In other words, there is no
local classical description (\emph{local hidden-variable theory}) that can
account for them, which are therefore \emph{purely quantum} and descend
directly from the \emph{non-locality} behavior of quantum mechanics (marked
by the violation of the\emph{\ Bell inequalities}). Exactly exploiting such
a kind of quantum correlations, it is possible to implement a purely \emph{%
quantum} teleportation, not reproducible by any classical strategies. Ref.~
\cite{Horodecki5} proved that every two-qubit state which violates some
\emph{generalized Bell inequality}, (representing, therefore, a channel with
non-local quantum correlations), allows to implement a teleportation with
fidelity $F>2/3$ (i.e., purely quantum). Consider, for instance, the \emph{%
Werner mixture }\cite{Werner}
\begin{equation}
\rho _{AB}(x)=x\left| \Psi _{AB}^{-}\right\rangle \left\langle \Psi
_{AB}^{-}\right| +(1-x)\frac{I}{4}~,  \label{Werner}
\end{equation}
where $I$ is the identity operator, $\left| \Psi _{AB}^{-}\right\rangle $ is
the EPR state of Eq.~(\ref{EPR_state_qubits}), and $0\leq x\leq 1$ is a
parameter quantifying the entanglement \cite{relation}. One can prove \cite
{Peres} that $\rho _{AB}(x)$ is entangled for $1/3\leq x\leq 1$, while the
Bell inequality is violated in the narrower region $1/\sqrt{2}\leq x\leq 1$.
Here, non-local quantum correlations between the subsystem $A$ and $B$
enable to implement a purely quantum teleportation, i.e., with fidelity $%
F>2/3$. In particular for $x=1$, $\rho _{AB}(1)$ coincides with the EPR
state $|\Psi _{AB}^{-}\rangle $ which is a maximally entangled state and
allows to perform an ideal quantum teleportation ($F=1$).

All the previous considerations concerned the world of qubits, which are
quantum systems described by a two-dimensional Hilbert space. In this paper,
instead, we consider quantum systems described by a Hilbert space with
infinite dimension. This is the case of the quantum oscillators or \emph{%
modes}, which can be, e.g., the radiation modes of the electromagnetic field
or the vibrating modes of a mechanical oscillator. Thanks to the infinite
dimensionality of their Hilbert space, these systems can be described by
observables, like position and momentum, which have a continuous spectrum of
eigenvalues. When we use these continuous variables to characterize their
state, we refer to these systems as continuous variable (CV) systems, and
the usage of such variables in order to encode, process and transmit both
classical and quantum information, gives rise to the CV quantum information
and computation theory \cite{cvbook}. In particular, we will review quantum
teleportation for CV systems. In Sec.~\ref{CV_systems} we will introduce the
necessary mathematical formalism to treat CV systems and Gaussian states.
Then, in Sec.~\ref{CV_entanglement} we will treat quantum entanglement for
CV systems, and in Sec.~\ref{CV_teleportation} we will describe quantum
teleportation in the CV framework together with its connection to CV quantum
entanglement. In the subsequent Sec.~\ref{CV_networks} we will extend CV
quantum teleportation from two to more users (CV quantum teleportation
networks)\ and we will study possible protocols in the case of three
parties. Finally, Sec.~\ref{Conclusion} is for conclusion.

\section{Continuous variable systems and Gaussian states\label{CV_systems}}

Consider $N$ quantum oscillators labelled by $k$, to which $N$ Hilbert
spaces $\mathcal{H}_{k}$ are associated. These modes are fully described by
ladder operators $\hat{a}_{k},\hat{a}_{k}^{\dagger }\in \mathcal{H}_{k}$
satisfying the bosonic commutation relations $[\hat{a}_{k},\hat{a}%
_{k^{\prime }}^{\dagger }]=\delta _{kk^{\prime }}$ and $[\hat{a}_{k},\hat{a}%
_{k^{\prime }}]=[\hat{a}_{k}^{\dagger },\hat{a}_{k^{\prime }}^{\dagger }]=0$%
. Equivalently, they can be fully described by dimensionless conjugate
quadratures \cite{Loock,QOptics}
\begin{equation}
\hat{x}_{k}\equiv (\hat{a}_{k}+\hat{a}_{k}^{\dagger })/\sqrt{2},~\hat{p}%
_{k}\equiv i(\hat{a}_{k}-\hat{a}_{k}^{\dagger })/\sqrt{2},  \label{P_k}
\end{equation}
satisfying the canonical commutation relations (CCR)
\begin{equation}
\lbrack \hat{x}_{k},\hat{p}_{k^{\prime }}]=i\delta _{kk^{\prime }},~~[\hat{x}%
_{k},\hat{x}_{k^{\prime }}]=[\hat{p}_{k},\hat{p}_{k^{\prime }}]=0.
\end{equation}
Introducing the vector of quadratures $\hat{\zeta}\equiv (\hat{x}_{1},\hat{p}%
_{1},...,\hat{x}_{N},\hat{p}_{N})^{T}$, the CCR be cast in the compact form
\begin{equation}
\lbrack \hat{\zeta}_{l},\hat{\zeta}_{m}]=i\mathbf{J}%
_{lm}^{(N)},~(l,m=1,...,2N)  \label{CCR}
\end{equation}
where
\begin{equation}
\mathbf{J}^{(N)}\equiv \bigoplus_{k=1}^{N}\mathbf{J},~\mathbf{J}\equiv
\left(
\begin{array}{cc}
0 & 1 \\
-1 & 0
\end{array}
\right) ,  \label{J_1mode}
\end{equation}
and $\oplus $\ denotes the direct sum. An arbitrary state $\rho $ of the $N$
modes is equivalently described by its symmetrically-ordered (or Wigner)
characteristic function $\Phi (\lambda )\equiv \mathrm{Tr}[\rho \hat{D}%
(\lambda )]$, where $\hat{D}(\lambda )=\exp (-i\hat{\zeta}^{T}\lambda )$ is
the Weyl operator and $\lambda \in \mathbb{R}^{2N}$. In particular, the
state $\rho $ is said to be Gaussian if the corresponding characteristic
function is Gaussian, i.e.,
\begin{equation}
\Phi (\lambda )=\exp [-\lambda ^{T}\frac{\mathbf{V}}{2}\lambda
+id^{T}\lambda ].  \label{Gaussian_CH_function}
\end{equation}
In such a case, the state $\rho $ is fully characterized by its displacement
$d\equiv \langle \hat{\zeta}\rangle $, and its correlation matrix (CM) $%
\mathbf{V}$, whose generic element is defined as
\begin{equation}
\mathbf{V}_{lm}\equiv \langle \Delta \hat{\zeta}_{l}\Delta \hat{\zeta}%
_{m}+\Delta \hat{\zeta}_{m}\Delta \hat{\zeta}_{l}\rangle /2  \label{CMdef}
\end{equation}
where $\Delta \hat{\zeta}_{l}\equiv \hat{\zeta}_{l}-\langle \hat{\zeta}%
_{l}\rangle $. All the information about the quantum \emph{and/or} classical
correlations among the oscillators is contained in the CM, which is a $%
2N\times 2N$, real and symmetric matrix satisfying the \emph{bona fide}
condition
\begin{equation}
\mathbf{V}+\frac{i}{2}\mathbf{J}^{(N)}\geq 0,  \label{CM_indeterm}
\end{equation}
which corresponds to the uncertainty principle \cite{CCRcompact}. By
applying the Fourier transform to Eq.~(\ref{Gaussian_CH_function}) one gets
the Wigner function of the arbitrary Gaussian state $\rho $ which is again a
Gaussian function and it is given by
\begin{equation}
W(\zeta )=\frac{\exp (-d^{T}\mathbf{V}^{-1}d/4)}{(2\pi )^{N}\sqrt{\det
\mathbf{V}}}\exp \left( -\zeta ^{T}\frac{\mathbf{V}^{-1}}{2}\zeta +\frac{%
d^{T}\mathbf{V}^{-1}}{\sqrt{2}}\zeta \right) .  \label{Wigner_Gaussiano}
\end{equation}
In Eq.~(\ref{Wigner_Gaussiano}), the variables $\zeta \equiv
(x_{1},p_{1},...,x_{N},p_{N})^{T}$, conjugate of $\lambda $, are the
continuous variables corresponding to the quadratures $\hat{\zeta}$, and
define the $2N-$dimensional phase space $\mathcal{K}$ of the system. These $%
2N$ real variables can be arranged\ into $N$ complex amplitudes
\begin{equation}
\alpha _{k}\equiv (x_{k}+ip_{k})/\sqrt{2},
\end{equation}
and the Wigner function of Eq.~(\ref{Wigner_Gaussiano}) can be expressed in
terms of the vector $\alpha \equiv (\alpha _{1},...,\alpha _{N})^{T}$ and
denoted as $W(\alpha )$.

Note that, since a Gaussian state of $N$ modes has a one-to-one
correspondence with its first and second moments, it requires only $%
2N^{2}+3N $ real parameters for its full description, which is polynomial
rather than exponential in $N$ \cite{IntroGauss}. This mathematical
tractability together with their experimental commonness have made Gaussian
states the object of an intensive study during the last years.

\subsection{Symplectic transformations}

The most general real \emph{linear} transformation $\mathbf{M}$\ over the
quadratures $\hat{\zeta}$,
\begin{equation}
\mathbf{M}:\hat{\zeta}\longrightarrow \hat{\zeta}^{\prime }\equiv \mathbf{M}%
\hat{\zeta},  \label{Linear_map}
\end{equation}
must preserve the CCR (\ref{CCR}) in order to be a physical operation. This
happens when the linear map~(\ref{Linear_map}) is described by a $2N\times
2N $, real matrix $\mathbf{M}$ [to be denoted as $\mathbf{M}\in \mathcal{M}%
(2N,\mathbb{R})$], which satisfies the condition
\begin{equation}
\mathbf{MJ}^{(N)}\mathbf{M}^{T}=\mathbf{J}^{(N)}~.  \label{Sympl_cond}
\end{equation}
The set of all matrices $\mathbf{M}\in \mathcal{M}(2N,\mathbb{R})$
satisfying Eq.~(\ref{Sympl_cond}) forms the so-called \emph{real symplectic
group} $\mathcal{S}_{p}(2N,\mathbb{R})$, whose elements are called \emph{%
symplectic} or \emph{canonical transformations}. An arbitrary symplectic
transformation~(\ref{Linear_map}) of the quadratures $\hat{\zeta}$ acts on
the corresponding continuous variables $\zeta $ of the phase space $\mathcal{%
K}$ as $\zeta \longrightarrow \zeta ^{\prime }\equiv \mathbf{M}\zeta $.
Therefore, it transforms the Wigner function as a scalar
\begin{equation}
W(\zeta )\longrightarrow W(\mathbf{M}^{-1}\zeta ^{\prime })~,
\end{equation}
and corresponds to a congruence at the level of the CM
\begin{equation}
\mathbf{V}\longrightarrow \mathbf{MVM}^{T}~.  \label{CM_congruence}
\end{equation}
Eq.~(\ref{CM_congruence}) represents the relevant phase-space transformation
when only the second moments are object of investigation. An arbitrary
symplectic transformation $\mathbf{M}$ acting on the phase space $\mathcal{K}
$ corresponds to a unitary operator $\hat{U}$ acting on the Hilbert space $%
\mathcal{H}$, which transforms the state according to the law $\rho
\longrightarrow \hat{U}\rho \hat{U}^{\dagger }$. The corresponding unitary
operator $\hat{U}=\hat{U}(\mathbf{M})$\ can be determined via the relation
\begin{equation}
\hat{U}^{\dagger }\hat{\zeta}\hat{U}=\mathbf{M}\hat{\zeta}.  \label{LUBO}
\end{equation}
Those particular unitary operators defined by Eq.~(\ref{LUBO}) are called
\emph{linear unitary Bogoliubov operations} (LUBOs) and, since they preserve
the Gaussian character of a density matrix, they are also called \emph{%
Gaussian unitary transformations} \cite{IntroGauss}. Symplectic
transformations of the form $\mathbf{M}=\bigoplus_{k=1}^{N}\mathbf{M}_{k}$
are \emph{local}, and correspond to \emph{local} linear unitary Bogoliubov
operations (LLUBOs) having the form $\hat{U}=\bigotimes_{k=1}^{N}\hat{U}_{k}$
with $\hat{U}_{k}$ acting only on $\mathcal{H}_{k}$.

Due to a theorem by Williamson \cite{Will}, the CM of a $N-$mode Gaussian
state can always be written as
\begin{equation}
\mathbf{V}=\mathbf{M~}%
\mbox{
       \boldmath{
           \small{\!\!\!\!\!$\nu$}}}\mathbf{M}^{T}  \label{Diag_sympl}
\end{equation}
where $\mathbf{M}\in \mathcal{S}_{p}(2N,\mathbb{R})$ and
\begin{equation}
\mbox{
       \boldmath{
           \small{\!\!\!\!\!$\nu $}}}=\left(
\begin{array}{ccccc}
\nu _{1} &  &  &  &  \\
& \nu _{1} &  &  &  \\
&  & \ddots &  &  \\
&  &  & \nu _{N} &  \\
&  &  &  & \nu _{N}
\end{array}
\right) \mathrm{.}  \label{Sympl_Spect}
\end{equation}
The $N$ quantities $\nu _{k}$'s in Eq.~(\ref{Sympl_Spect}) represent the
\emph{symplectic eigenvalues} of the CM $\mathbf{V}$\ and they are clearly
invariant under symplectic transformations. In the Hilbert space, the dual
formulation of Eq.~(\ref{Diag_sympl}) reads
\begin{equation}
\rho =\hat{U}\left( \bigotimes_{k=1}^{N}\bar{\rho}_{k}\right) \hat{U}%
^{\dagger }~,
\end{equation}
where $\bar{\rho}_{k}$ is a thermal state \cite{Gardiner} having a
mean excitation number $\bar{n}_{k}$\ such that $\nu
_{k}=\bar{n}_{k}+1/2$. The symplectic eigenvalues $\nu _{k}$'s
encode essential information on the Gaussian state, and provide
powerful ways to express its fundamental properties. For instance,
the condition $\nu _{k}=1/2$ characterizes pure states, as it can
be shown by the Von Neumann entropy which is equal to \cite
{Holevo}
\begin{equation}
H(\rho )=\sum_{k=1}^{N}g(\nu _{k})~,
\end{equation}
where
\begin{equation}
g(x)\equiv (x+1/2)\log (x+1/2)-(x-1/2)\log (x-1/2).
\end{equation}
In particular, the uncertainty principle of Eq.~(\ref{CM_indeterm}) can be
cast in the following equivalent form
\begin{equation}
\mathbf{V}+\frac{i}{2}\mathbf{J}^{(N)}\geq 0~\Leftrightarrow ~\nu _{k}\geq
\frac{1}{2}~,  \label{equiv_Heis}
\end{equation}
which is saturated only by pure Gaussian states.

Consider now the particular case of Gaussian states of only \emph{two}
modes. In this case, it is convenient to write the CM in the blockform
\begin{equation}
\mathbf{V}=\left(
\begin{array}{cc}
\mathbf{A} & \mathbf{C} \\
\mathbf{C}^{T} & \mathbf{B}
\end{array}
\right) ~,  \label{V_blocks_2modi}
\end{equation}
where $\mathbf{A},\mathbf{B},\mathbf{C}\in \mathcal{M}(2,\mathbb{R})$ and $%
\mathbf{A}=\mathbf{A}^{T},~\mathbf{B}=\mathbf{B}^{T}$. For bipartite
Gaussian states, the two symplectic eigenvalues of the CM are proved \cite
{Alessio} to be equal to
\begin{equation}
\nu _{\pm }=\sqrt{\frac{\Delta (\mathbf{V})\pm \sqrt{\Delta (\mathbf{V}%
)^{2}-4\det \mathbf{V}}}{2}},  \label{symple}
\end{equation}
where
\begin{equation}
\Delta (\mathbf{V})\equiv \det \mathbf{A}+\det \mathbf{B}+2\det \mathbf{C~.}
\label{seraliano}
\end{equation}
Here, the quantities $\Delta (\mathbf{V})$\ and $\det \mathbf{V}$ are
invariant under (global) symplectic transformations. Furthermore, suitable
local symplectic transformations $\mathbf{M}_{1}\oplus \mathbf{M}_{2}$
allows to transform the CM of Eq.~(\ref{V_blocks_2modi}) into its \emph{%
normal form} or \emph{standard form I} \cite{SimonPRL}

\begin{equation}
\mathbf{V}^{\text{I}}=\left(
\begin{array}{cccc}
a &  & c &  \\
& a &  & c^{\prime } \\
c &  & b &  \\
& c^{\prime } &  & b
\end{array}
\right) \equiv \mathbf{V}^{\text{I}}(a,b,c,c^{\prime })~,
\label{normal_FORM}
\end{equation}
where $a^{2}=\det \mathbf{A},~b^{2}=\det \mathbf{B},~cc^{\prime }=\det
\mathbf{C}$ and $\det \mathbf{V}^{\text{I}}=\det \mathbf{V}$ are four local
symplectic invariants of the original CM.

\section{Continuous variable entanglement\label{CV_entanglement}}

\subsection{Peres-Horodecki separability criterion for continuous variable
systems}

Consider two quantum systems labelled by letter $A$ (referring to Alice) and
$B$ (referring to Bob), which correspond to two Hilbert spaces $\mathcal{H}%
_{A}$ and $\mathcal{H}_{B}$, respectively. An arbitrary state of these two
systems $\rho _{AB}\in \mathcal{H}_{AB}=\mathcal{H}_{A}\otimes \mathcal{H}%
_{B}$ is called \emph{separable} when it can be written as a convex sum of
tensor products of single-party states \cite{Werner}, i.e.,
\begin{equation}
\rho _{AB}=\sum_{k}p_{k}\rho _{A}^{k}\otimes \rho _{B}^{k}~,~~~p_{k}\geq
0~,~\sum_{k}p_{k}=1~,  \label{separable}
\end{equation}
with $\rho _{A}^{k}\in \mathcal{H}_{A}$ and $\rho _{B}^{k}\in \mathcal{H}%
_{B} $. In such a case the bipartite state can be prepared via local
operations and classical communications (LOCCs) acting on two uncorrelated
subsystems $A $ and $B$. On the contrary, $\rho _{AB}$ is said \emph{%
entangled} when Eq.~(\ref{separable}) does not hold.

In the case of finite dimensional Hilbert spaces, i.e., $\dim \mathcal{H}%
_{A}=d_{A}$ and $\dim \mathcal{H}_{B}=d_{B}$, a simple criterion to test
separability was derived by Peres \cite{Peres} who resorted to the so-called
\emph{partial transpose} (PT) operation. Introducing an orthonormal basis in
the Hilbert space $\mathcal{H}_{AB}$, the arbitrary state of the bipartite
system $A+B$ is described by a density matrix $(\rho _{AB})_{m\mu ,n\nu }$,
where Latin indices refer to the first subsystem ($1\leq m,n\leq d_{A}$) and
Greek indices to the second one ($1\leq \mu ,\nu \leq d_{B}$). Now the PT
operation can be defined as an inversion of the Greek indices
\begin{equation}
\mathrm{PT}:~(\rho _{AB})_{m\mu ,n\nu }\rightarrow (\rho _{AB})_{m\nu ,n\mu
}\equiv \lbrack \mathrm{PT}(\rho _{AB})]_{m\mu ,n\nu }~.  \label{PT}
\end{equation}
In Eq.~(\ref{PT}) we ask if the operator $\mathrm{PT}(\rho _{AB})$ is still
a density operator, i.e., if $\mathrm{Tr}[\mathrm{PT}(\rho _{AB})]=1$ and $%
\mathrm{PT}(\rho _{AB})\geq 0$. Since the PT operation preserves the
diagonal elements of the density matrix, the genuineness of the final state
simply corresponds to its positivity. By definition, those density operators
$\rho _{AB}$, for which $\mathrm{PT}(\rho _{AB})\geq 0$ holds, are called
\emph{positive partial transpose }(PPT) states. Peres showed \cite{Peres}
that the PPT property is a necessary condition for separability
\begin{equation}
d_{A}\times d_{B}~\text{\emph{system}}~~:~~\rho
_{AB}~separable\Longrightarrow \mathrm{PT}(\rho _{AB})\geq 0~,
\label{CN_Peres}
\end{equation}
and, in the particular case of two qubits, it is also sufficient
\begin{equation}
2\times 2~\text{\emph{system}}~~:~~\rho _{AB}~separable\Longleftrightarrow
\mathrm{PT}(\rho _{AB})\geq 0~.  \label{PPTqubits}
\end{equation}

Later on, R. Simon \cite{SimonPRL} showed how to extend the PT operation (%
\ref{PT}) and the Peres criterion (\ref{CN_Peres}-\ref{PPTqubits}) to the
case of CV bipartite states, i.e., for bipartite states of two modes $A$ and
$B$. Let us introduce the phase space representation of a CV bipartite state
$\rho _{AB}$ via its Wigner function $W(\zeta ),$ where $\zeta ^{T}\equiv
(x_{A},p_{A},x_{B},p_{B})$. In phase space, transposition is defined as
``time reversal'' (or mirror reflection) which is given by a change of sign
of momentum operators. Partial transposition is therefore a ``local time
reversal'' which inverts the momentum of only one subsystem. Thus, the PT
operation on the state $\rho _{AB}$ in $\mathcal{H}_{AB}$%
\begin{equation}
\mathrm{PT}:\rho _{AB}\longrightarrow \mathrm{PT}(\rho _{AB})
\end{equation}
is equivalent to the following partial mirror reflection of the Wigner
function $W(\zeta )$ in phase space $\mathcal{K}$%
\begin{equation}
W(\zeta )\longrightarrow W(\mathbf{\Lambda }\zeta )~,  \label{Partial_refle}
\end{equation}
where $\mathbf{\Lambda }\equiv \mathbf{I}\oplus \mathbf{Z}$, with
\begin{equation}
\mathbf{Z}\equiv \left(
\begin{array}{cc}
1 &  \\
& -1
\end{array}
\right) ,  \label{reflection}
\end{equation}
and $\mathbf{I}$ the $2\times 2$ identity matrix. Consequently, the CM\ of
the state is transformed according to the law
\begin{equation}
\mathbf{V}\longrightarrow \mathbf{\Lambda V\Lambda ~.}
\end{equation}
For this reason, we can extend the Peres criterion~(\ref{CN_Peres}) to CV
bipartite states as
\begin{equation}
\rho _{AB}~separable\Longrightarrow \mathbf{\Lambda V\Lambda }\text{ is a
\emph{bona fide} CM,}  \label{CN_Peres_CV}
\end{equation}
where \emph{bona fide} means ``corresponding to a physical state'' and it is
expressed by condition of Eq.~(\ref{CM_indeterm}) at the level of the second
moments. Therefore, we can write the necessary condition
\begin{equation}
\rho _{AB}~separable\Longrightarrow \mathbf{\Lambda V\Lambda }+\frac{i}{2}%
\mathbf{J}^{(2)}\geq 0~,  \label{CN_Peres_CV_CM}
\end{equation}
where the right-hand of Eq.~(\ref{CN_Peres_CV_CM}) is also known as PPT
property for the CM $\mathbf{V}$, by analogy with Eq.~(\ref{CN_Peres}). More
strongly, one can prove \cite{SimonPRL} the following theorem in the case of
Gaussian states

\begin{description}
\item[\textbf{Theorem.}]  Consider an arbitrary bipartite Gaussian state $%
\rho _{AB}$. Then
\begin{equation}
\rho _{AB}~separable\Leftrightarrow \mathbf{\Lambda V\Lambda }+\frac{i}{2}%
\mathbf{J}^{(2)}\geq 0~(\mathbf{V}\text{ is PPT}).  \label{CNS_Gauss}
\end{equation}
\end{description}

\noindent Note that, in the right hand of the previous Eq.~(\ref{CNS_Gauss}%
), CM $\mathbf{V}$ must also implicitly satisfy the bona-fide condition of
Eq.~(\ref{CM_indeterm}). This remark is important when the entanglement
properties are studied through the second moments and completely arbitrary
CM's are considered in the investigation, as, for instance, in deriving
computable criteria for separability \cite{Commento}.

The study of CV entanglement acquires an elegant form by resorting to the
formalism of symplectic invariants. Consider an arbitrary Gaussian state $%
\rho _{AB}$ whose CM $\mathbf{V}$ is put in the blockform of Eq.~(\ref
{V_blocks_2modi}). One can then verify that the PT\ transformation $\mathbf{%
\Lambda }$ reduces to the sign flipping $\det
\mathbf{C}\longrightarrow -\det \mathbf{C}$ at the level of the
local symplectic invariants. For this reason, the two symplectic
eigenvalues $\tilde{\nu}_{\pm }$ of the partially transposed CM
$\mathbf{\tilde{V}\equiv \Lambda V\Lambda }$ satisfy Eq.~(\ref
{symple}) except for the replacement
\begin{equation}
\Delta (\mathbf{V})\longrightarrow \tilde{\Delta}(\mathbf{V})\equiv \det
\mathbf{A}+\det \mathbf{B}-2\det \mathbf{C~.}
\end{equation}
If we now apply Eq.~(\ref{equiv_Heis}) to the equivalence of Eq.~(\ref
{CNS_Gauss}), we can conclude the following

\begin{description}
\item[\textbf{Theorem.}]  Consider an arbitrary bipartite Gaussian state $%
\rho _{AB}$. Then
\begin{equation}
\rho _{AB}~separable\Leftrightarrow \tilde{\nu}_{-}\geq 1/2~.
\label{theorem_symple}
\end{equation}
\end{description}

In general, one is able to provide not only a test for bipartite
entanglement as before, but also computable measures of entanglement, called
\emph{entanglement monotones} \cite{VidMono}, whose fundamental property is
to be monotonic under LOCCs \cite{HOROmono}. This is the case of the
negativity and logarithmic negativity which are defined for discrete
variable states and suitably extended to Gaussian states of CV systems \cite
{VidalW}. Consider an arbitrary bipartite state $\rho _{AB}$ of a $%
d_{A}\times d_{B}$ system, its negativity is defined as
\begin{equation}
\mathcal{N}(\rho _{AB})=\frac{\left\| \mathrm{PT}(\rho _{AB})\right\| _{1}-1%
}{2},
\end{equation}
where $\left\| \rho \right\| _{1}\equiv \mathrm{Tr}\left| \rho \right| $ is
the trace norm (Schatten 1-norm). One can verify that $\mathcal{N}(\rho
_{AB})$ is equal to the modulus of the sum $\left| \sum_{i}\lambda
_{i}\right| $ of the negative eigenvalues $\lambda _{i}$ of $\mathrm{PT}%
(\rho _{AB})$. In this sense, the negativity quantifies the extent to which $%
\mathrm{PT}(\rho _{AB})$ fails to be positive and, intuitively, represents a
measure of the bipartite entanglement within $\rho _{AB}$. Similarly, the
logarithmic negativity is defined as $E_{\mathcal{N}}(\rho _{AB})=\log
\left\| \mathrm{PT}(\rho _{AB})\right\| _{1}$ and has the remarkable
property to be additive under tensor product. These measures can be readily
extended to bipartite Gaussian states by interpreting the PT operation as
partial mirror reflection as before. In such a case, one can prove \cite
{Salerno1} that
\begin{equation}
E_{\mathcal{N}}(\rho _{AB})=\max [0,-\log 2\tilde{\nu}_{-}]\text{,}
\end{equation}
where $\tilde{\nu}_{-}$ is the minimum PT-symplectic eigenvalue of $\rho
_{AB}$. Thanks to its monotonic relation with the logarithmic negativity,
the minimum PT-symplectic eigenvalue $\tilde{\nu}_{-}$\ can be adopted as
entanglement monotone for bipartite Gaussian states.

\subsection{Bipartite entanglement and EPR correlations}

Besides R. Simon, also L.-M. Duan \textit{et al.}~\cite{DuanPRL}\ derived a
criterion for entanglement, which turns out to be very useful for
teleportation. As usual, consider two modes, $A$ and $B$, with quadrature
operators $\hat{x}_{A},\hat{p}_{A}$ and $\hat{x}_{B},\hat{p}_{B}$. From
these operators, we can define the following EPR-like operators
\begin{equation}
\hat{L}(q)\equiv |q|\hat{x}_{A}+q^{-1}\hat{x}_{B}~,~\hat{M}(q)\equiv |q|\hat{%
p}_{A}-q^{-1}\hat{p}_{B}~,  \label{EPR_LikeOp}
\end{equation}
where $q\in \mathbb{R}\setminus 0$. Note that for $q=-1$ we have the
standard EPR\ operators
\begin{equation}
\hat{X}_{-}\equiv \hat{x}_{A}-\hat{x}_{B}~,~\hat{P}_{+}\equiv \hat{p}_{A}+%
\hat{p}_{B}~.  \label{EPR_Operators}
\end{equation}
An arbitrary state $\rho _{AB}$\ of the two-mode system satisfies the
following necessary condition for separability
\begin{equation}
\rho _{AB}\text{~\emph{separable}}\Longrightarrow \left\langle \lbrack
\Delta \hat{L}(q)]^{2}\right\rangle +\left\langle [\Delta \hat{M}%
(q)]^{2}\right\rangle \geq q^{2}+q^{-2}~,~\forall q\in \mathbb{R}\setminus 0,
\label{Duan_sep}
\end{equation}
where $\langle \lbrack \Delta \hat{L}(q)]^{2}\rangle $ and $\langle \lbrack
\Delta \hat{M}(q)]^{2}\rangle $ denote the variances of $\hat{L}(q)$ and $%
\hat{M}(q)$, computed over the state $\rho _{AB}$. From (\ref{Duan_sep}) one
derives the following sufficient criterion for entanglement

\begin{description}
\item[\textbf{Criterion.}]  If $\exists q\in \mathbb{R}\setminus 0$ such
that
\begin{equation}
\left\langle \lbrack \Delta \hat{L}(q)]^{2}\right\rangle +\left\langle
[\Delta \hat{M}(q)]^{2}\right\rangle <q^{2}+q^{-2}~,  \label{Duan_criterium}
\end{equation}
then$~\rho _{AB}$~is entangled.\label{EPR_CHANNEL}
\end{description}

\noindent As we have already said, a bipartite Gaussian state $\rho _{AB}$
can be fully described by its displacement $d$ and CM$~\mathbf{V}$. Since
its separability properties do not vary under unitary operations, we may
cancel its displacement $d$ via local displacement operators, and reduce its
CM $\mathbf{V}$ to the \emph{normal form} $\mathbf{V}^{\text{I}%
}(a,b,c,c^{\prime })$ of Eq.~(\ref{normal_FORM}) via a sequence of local
symplectic transformations (local squeezings and rotations). The final state
has the same separability properties of the original one, but it is
associated to a simpler pair $d=0$ and$~\mathbf{V}^{\text{I}%
}(a,b,c,c^{\prime })$. Consider now those states having a CM of the type $%
\mathbf{V}^{\text{I}}(a,b,c,-c)$ whose peculiarity is to satisfy the
equality
\begin{equation}
\left\langle \Delta \hat{X}_{-}^{2}\right\rangle =\left\langle \Delta \hat{P}%
_{+}^{2}\right\rangle =a+b-2c\equiv \aleph ~.  \label{EPR_measure}
\end{equation}
For these states, one can use the quantity $\aleph \geq 0$ as a measure of
the correlations between the position and momentum observables of modes $A$
and $B$. In fact, here, $\aleph $ quantifies the two-mode squeezing in both
the EPR operators $\hat{X}_{-}$ and $\hat{P}_{+}$. For $\aleph <1$ we say,
by definition, that the two modes possess \emph{EPR\ correlations} in $\hat{X%
}_{-}$ and $\hat{P}_{+}$, and the corresponding state is an \emph{EPR channel%
}. Note that $\aleph <1$ implies that $\rho _{AB}$ is an entangled state
(from Eq.~(\ref{Duan_criterium}) with $q=-1$). But if $\aleph \geq 1$ one
cannot exclude the existence of another pair of EPR-like operators $\hat{L}%
(q)$ and $\hat{M}(q)$ which implies entanglement according to Eq.~(\ref
{Duan_criterium}). Thus, entanglement is only a necessary condition for the
existence of EPR\ correlations for a particular pair of conjugate
quadratures. If we take the limit $\aleph \longrightarrow 0$ in Eq.~(\ref
{EPR_measure}), we achieve the perfect EPR correlations
\begin{equation}
\hat{x}_{A}-\hat{x}_{B}=\hat{p}_{A}+\hat{p}_{B}=0~,  \label{Ideal_EPR}
\end{equation}
i.e., we achieve the ideal EPR\ state \cite{EPR}. This state represents the
CV\ version of the discrete EPR\ pair of Eq.~(\ref{EPR_state_qubits}). It is
a maximally entangled state ($E_{\mathcal{N}}=+\infty $) and, as we shall
see, it allows to perform an ideal quantum teleportation.

Consider now the \emph{two-mode squeezed vacuum }(TMSV) state or \emph{twin
beam} (TB) state, which is a bipartite Gaussian state having Wigner function
\begin{equation}
W_{TB}(\zeta )=\frac{1}{\pi ^{2}}\exp \left( -\zeta ^{T}\frac{\mathbf{V}%
(r)^{-1}}{2}\zeta \right) ,~\zeta ^{T}=(x_{A},p_{A},x_{B},p_{B})~,
\label{Wig_TB}
\end{equation}
where the CM
\begin{equation}
\mathbf{V}(r)=\frac{1}{2}\left(
\begin{array}{cc}
(\cosh 2r)\mathbf{I} & (\sinh 2r)\mathbf{Z} \\
(\sinh 2r)\mathbf{Z} & (\cosh 2r)\mathbf{I}
\end{array}
\right)  \label{CM_TB}
\end{equation}
depends on a squeezing factor $r$. Note that the CM of Eq.~(\ref{CM_TB}) is
a particular CM of the type $\mathbf{V}^{\text{I}}(a,b,c,-c)$, showing that
the twin beam is an example of EPR\ channel. One can easily verify that $%
\aleph =\exp (-2r)$ which shows that, for every $r>0$, one effectively has $%
\aleph <1$ corresponding to two-mode squeezing, i.e., EPR\ correlations. If
we now take the limit of infinite squeezing $r\longrightarrow +\infty $ in
Eq.~(\ref{Wig_TB}), we achieve
\begin{equation}
W_{TB}(\zeta )\longrightarrow W_{EPR}(\zeta )\varpropto \delta
(x_{A}-x_{B})\delta (p_{A}+p_{B})~,  \label{limite_EPR}
\end{equation}
i.e., we obtain the ideal EPR state \cite{EPR}. As we shall see, this
asymptotic state enables to implement a perfect CV quantum teleportation.
However, in any real experimental setup, one can only access finite
squeezing, and, therefore, exploit quantum channels as the TMSV state of
Eq.~(\ref{Wig_TB}).

\section{Continuous variable quantum teleportation\label{CV_QT}\label%
{CV_teleportation}}

Thanks to the elements developed in the previous sections, we are now ready
to discuss quantum teleportation within the continuous variable framework.
First of all, one can ask why the need of a CV\ quantum teleportation, since
the discrete one already exists. Some important facts can be quoted to
justify this further development of teleportation and they are mainly
connected with experimental advantages. In fact, one of the key-points of
the discrete variable protocol is the usage of a Bell measurement in order
to couple the input qubit with the quantum\ channel. Till now, in all the
real experimental setups, such a measurement has not been implemented in a
clear and precise way. In other words, the perfect Bell-state discrimination
does not seem to be achievable in a simple way, e.g., by linear passive
elements as beam splitters and photodetectors~\cite{BellB0}. On the
contrary, the CV version of the Bell measurement is realized via linear
passive optics and homodyne measurements, whose outcomes can be
discriminated with high precision (perfectly discriminated in asymptotic
sense).

Essentially, CV\ quantum teleportation is based on the same ideas of the
discrete case. The basic resource is always the same, i.e., (CV)
entanglement shared between Alice and Bob, which is a necessary condition in
order to have a pair of\ (CV) EPR-correlated observables to be exploited in
the protocol by the two parties. For sufficiently high entanglement, quantum
channel reveals non-local quantum correlations that enable to perform a
\emph{truly} quantum teleportation, i.e., a teleportation process with
fidelity $F>F_{class}$, where $F_{class}$\ is the fidelity of the best
classical strategy. In the CV\ case, $F_{class}=0$ for \emph{arbitrary}
input states (instead of $2/3$, as for the qubits), while, for the
particular task of teleporting coherent states, one has $%
F_{class}^{(coh)}=1/2$ \cite{F1su2}. Such a threshold has been proved to be
connected with the violation of a CHSH inequality in Ref.~\cite{ClassicF}.

The protocol can be shown by analogy with the one for qubits. As usual,
Alice mixes the unknown input state with the shared quantum channel via a
Bell measurement, which, in continuous variables, is realized by a beam
splitter and two homodyne detectors. In this way she sends part of
information on the state to Bob via the quantum channel. In order to
complete the process, Bob needs also to know the result of Alice's
measurement and this is exactly the part of classical information on the
state that Alice communicates to Bob through the classical channel (this
classical information is equal to two bits in the discrete case, while it is
given by two real variables in the CV case). Conditioned to this classical
information, Bob performs a suitable unitary operation on his part of
quantum channel (which is a Pauli operator in the discrete case, while it is
a displacement operator in CV) and correctly reconstructs the original input
state that was in Alice's hands. The net result of the teleportation is the
removal of the input state from Alice's side and its appearance in Bob's
side some time later (the time needed by a classical message to go from
Alice to Bob).

In the following we will first present the ideal situation of a perfect CV
quantum teleportation, which one has in the case of an ideal EPR pair as
quantum channel. Such analysis will be carried out in the Heisenberg picture
for its simplicity. Then, we will switch to a more real situation, where the
quantum channel is imperfect (given by an arbitrary two-mode state), and
this subsequent analysis will be carried out in the Schroedinger picture and
using the Wigner representation in phase space.

\subsection{Ideal CV quantum teleportation\label{Ideal_CV_QT}}

\begin{figure}[tbph]
\vspace{+0.2cm}
\par
\begin{center}
\includegraphics[width=0.99\textwidth]{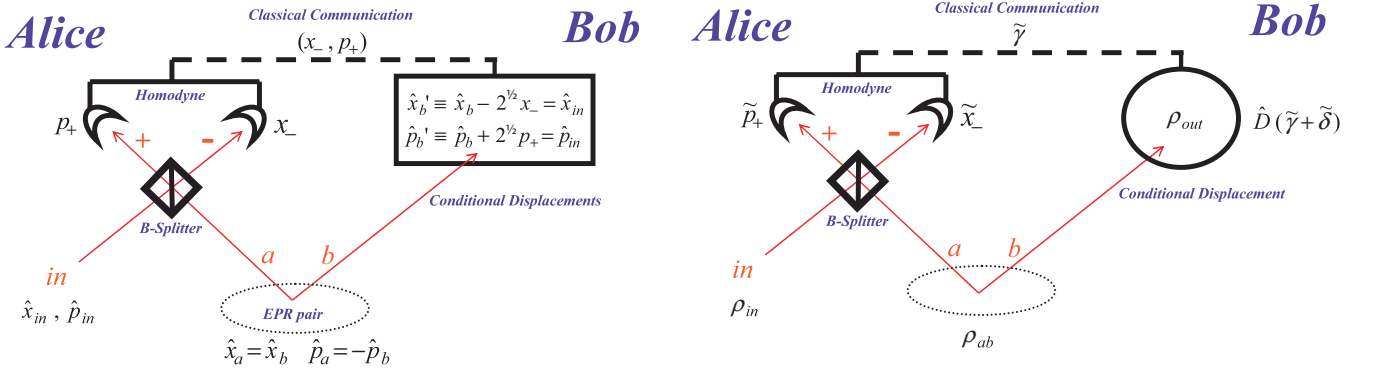}
\end{center}
\par
\vspace{-0.3cm} \caption{Ideal (left side) and real (right side)
quantum teleportation.} \label{CVTele}
\end{figure}

The proposal for a CV quantum teleportation was firstly made by Vaidman in
Ref.~\cite{vaidman} and then by Braunstein and Kimble in Ref.~\cite
{KimbleTele}. The ideal situation of a quantum channel given by an ideal
EPR\ pair can be easily shown in the Heisenberg picture. The protocol goes
as follows (see also Fig.~\ref{CVTele}, left side):

\begin{enumerate}
\item  \emph{Initial condition}. Alice and Bob share a quantum channel given
by an ideal EPR pair, i.e., two modes $a$ and $b$ with quadratures such that
\begin{equation}
\hat{x}_{a}-\hat{x}_{b}=\hat{p}_{a}+\hat{p}_{b}=0~.  \label{CVIDEAL1}
\end{equation}
Then, Alice has an unknown input state described by an input mode $in$ with
quadratures $\hat{x}_{in},\hat{p}_{in}$.

\item  \emph{Bell measurement}. Alice realizes a CV version of the Bell
measurement combining two subsequent operations on her modes

\begin{enumerate}
\item  \emph{Beam splitter mixing}. Alice mixes the input mode with her $a$
mode (part of the EPR pair) via a balanced (lossless) beam splitter, i.e.,
she performs the following bilinear transformation on their quadratures
\begin{equation}
\hat{x}_{\pm }=(\hat{x}_{a}\pm \hat{x}_{in})/\sqrt{2}~,~\hat{p}_{\pm }=(\hat{%
p}_{a}\pm \hat{p}_{in})/\sqrt{2}~,  \label{XP_BSplitter}
\end{equation}
where $\hat{x}_{\pm }$ and $\hat{p}_{\pm }$ are the quadratures of the
output modes ``$+$'' and ``$-$'' of the beam splitter.

\item  \emph{Homodyne detection}. Alice homodynes the output modes ``$\pm $%
''. In particular, she detects the quadratures $\hat{x}_{-}$ and $\hat{p}_{+}
$, i.e., she applies the projectors $\left| x\right\rangle \left\langle
x\right| $ to mode ``$-$'' and $\left| p\right\rangle \left\langle p\right| $
to mode ``$+$'' \cite{notaEPR}. Denoting with $(x_{-},p_{+})$ the two
outcomes, her measurement causes in Eq.~(\ref{XP_BSplitter}) the collapse
\begin{equation}
\hat{x}_{a}=\hat{x}_{in}+\sqrt{2}x_{-}~,~\hat{p}_{a}=-\hat{p}_{in}+\sqrt{2}%
p_{+}~.  \label{XP_collapse}
\end{equation}
Due to the EPR property~(\ref{CVIDEAL1}), Bob's quadratures, $\hat{x}_{b}$
and $\hat{p}_{b}$, are instantaneously projected according to (\ref
{XP_collapse}), i.e., we have
\begin{equation}
\hat{x}_{b}=\hat{x}_{in}+\sqrt{2}x_{-}~,~\hat{p}_{b}=\hat{p}_{in}-\sqrt{2}%
p_{+}~.  \label{Bob_collapse}
\end{equation}
\end{enumerate}

\item  \emph{Classical communication}. At this point, Alice communicates her
measurement result $(x_{-},p_{+})$ to Bob through a classical channel.

\item  \emph{Conditional displacement}. Bob uses this transmitted classical
information $(x_{-},p_{+})$ to perform a suitable conditional displacement
on his own mode $b$ which allows to complete the teleportation process
\begin{equation}
\begin{array}{l}
\hat{x}_{b}\longrightarrow \hat{x}_{b}^{\prime }\equiv \hat{x}_{b}-\sqrt{2}%
x_{-}=\hat{x}_{in}~, \\
\hat{p}_{b}\longrightarrow \hat{p}_{b}^{\prime }\equiv \hat{p}_{b}+\sqrt{2}%
p_{+}=\hat{p}_{in}~.
\end{array}
\label{Bob_displ}
\end{equation}
\end{enumerate}

\noindent In fact, according to Eq.~(\ref{Bob_displ}), mode $b$ is finally
described by a pair of conjugate quadratures ($\hat{x}_{b}^{\prime }$ and $%
\hat{p}_{b}^{\prime }$) which are exactly the ones of the input ($\hat{x}%
_{in}$ and $\hat{p}_{in}$). In the Schroedinger picture, this is equivalent
to teleport the quantum state from the input mode $in$, at Alice's side, to
the output mode $b$, at Bob's side, with fidelity $F=1$.

\subsection{Real CV quantum teleportation\label{REAL_CV_QT}}

As we have already said in the introduction of Sec.~\ref{CV_QT}, the
previous situation, based on an ideal EPR pair, can only be considered as a
limit of infinite squeezing of a real teleportation process, where the
quantum channel has a finite amount of squeezing. Actually, also the
homodyne measurement cannot be infinitely precise, and this detection
inefficiency can be explicitly taken into account \cite{DABELL}. Here, we
will ignore this measurement imperfection and we will focus our attention on
the more fundamental point concerning the reality of the channel. In
particular, our analysis is carried out for quantum channels given by
two-mode Gaussian states.

In this section we first show CV quantum teleportation through an\emph{\
arbitrary} two-mode state, adopting the Schroedinger picture and, in
particular, the phase space Wigner representation \cite{Chiz,FIU,JMO}. We
repeat the previous four-step analysis of the protocol in this general
situation and using this different formalism \cite{Chiz}. Then, we will
consider the case of a Gaussian two-mode state, and we will obtain a very
simple and useful formula for the fidelity \cite{FIU,JMO}. Such a case
comprises the Gaussian EPR\ channel (see Sec.~\ref{EPR_CHANNEL}) and, in
particular, the TMSV which is the real analog of the ideal EPR pair. The
protocol goes as follows (see also Fig.~\ref{CVTele}, right side):

\begin{enumerate}
\item  \emph{Initial condition}. Alice and Bob share a quantum channel given
by an arbitrary state $\rho _{ab}$ of two modes $a$ and $b$. Such a state
can be fully characterized by its Wigner function $W^{ch}(\alpha ,\beta )$,
where the complex amplitudes $\alpha =(x_{a}+ip_{a})/\sqrt{2}$ and $\beta
=(x_{b}+ip_{b})/\sqrt{2}$ refer to mode $a$ and $b$, respectively. Then,
Alice has an unknown input state $\rho _{in}$\ of a single mode $in$, whose
complex amplitude is $\gamma =(x_{in}+ip_{in})/\sqrt{2}$ and Wigner function
$W^{in}(\gamma )$. The total initial state is given by the tensor product $%
\rho _{in}\otimes \rho _{ab}$, and therefore it is described by the product
\begin{equation}
W(\gamma ,\alpha ,\beta )=W^{in}(\gamma )W^{ch}(\alpha ,\beta )~.
\label{Total_before}
\end{equation}

\item  \emph{Bell measurement}. Alice realizes a CV version of the Bell
measurement combining two subsequent operations on her modes

\begin{enumerate}
\item  \emph{Beam splitter mixing}. Alice mixes the input mode with her $a$
mode via a balanced (lossless) beam splitter, i.e., she performs the
following bilinear transformation on the complex amplitudes
\begin{equation}
\nu _{\pm }=(\alpha \pm \gamma )/\sqrt{2}~,  \label{BS_ComplexVariable}
\end{equation}
where $\nu _{\pm }\equiv (x_{\pm }+ip_{\pm })/\sqrt{2}$ are the complex
amplitudes of the output modes $\pm $. The total state just after the beam
splitter is given by
\begin{equation}
W(\nu _{-},\nu _{+},\beta )=W^{in}\left( \frac{\nu _{+}-\nu _{-}}{\sqrt{2}}%
\right) W^{ch}\left( \frac{\nu _{+}+\nu _{-}}{\sqrt{2}},\beta \right) ~,
\label{Total_after}
\end{equation}
which is derived from Eq.~(\ref{Total_before}) by using the inverse
transformations of Eq.~(\ref{BS_ComplexVariable}).

\item  \emph{Homodyne detection}. Alice detects the two quadratures $\hat{x}%
_{-}$ and $\hat{p}_{+}$, and the corresponding outcomes $(\tilde{x}_{-},%
\tilde{p}_{+})$ can be expressed by a single complex number
\begin{equation}
\tilde{\gamma}\equiv -\tilde{x}_{-}+i\tilde{p}_{+}~,  \label{Outcome_C}
\end{equation}
having probability $\mathcal{P}(\tilde{\gamma})$. Due to her measurement,
Alice remotely creates at Bob's station (mode $b$) a conditioned state whose
Wigner function is equal to
\begin{equation}
W(\beta |\tilde{\gamma})=\mathcal{P}(\tilde{\gamma})^{-1}\int d^{2}\gamma
~W^{in}(\gamma )W^{ch}(\gamma ^{\ast }-\tilde{\gamma}^{\ast },\beta )~.
\label{Cond_Bob_Wig2bis}
\end{equation}
\end{enumerate}

\item  \emph{Classical communication}. Alice communicates her measurement
result $\tilde{\gamma}$ to Bob through a classical channel. After that Bob
has received this classical information, the state of his mode $b$ is
described by the Wigner function of Eq.~(\ref{Cond_Bob_Wig2bis}).

\item  \emph{Conditional displacement}. Bob uses the received classical
information $\tilde{\gamma}$ to perform a suitable conditional displacement
on his own mode $b$ which aims to complete the teleportation process. What
is now the ``right'' displacement? For a \emph{zero-}displaced two-mode
state, like the EPR pair of Eq.~(\ref{CVIDEAL1}), the right displacement is
given by Eq.~(\ref{Bob_displ}), which is equivalent to the displacement
\begin{equation}
\beta \longrightarrow \beta ^{\prime }=\beta +\tilde{\gamma}~.
\label{Displ_Complex_Bob}
\end{equation}
The quantity $\tilde{\gamma}$ in Eq.~(\ref{Displ_Complex_Bob}) is an
extra-shift with respect to the original input state which is naturally
generated by the process and, in particular, by the Bell measurement. In
general, for a quantum channel given by a two-mode state with a \emph{%
nonzero }displacement, Bob's displacement must be composed by two terms.
Besides $\tilde{\gamma}$, accounting for detection, there is another complex
number $\tilde{\delta}\equiv \tilde{\delta}^{R}+i\tilde{\delta}^{I}$ which
balances the shift inherited by the channel. Thus, the correct displacement
has the form
\begin{equation}
\beta \longrightarrow \beta ^{\prime }=\beta +\tilde{\gamma}+\tilde{\delta}~,
\label{Displ_correct}
\end{equation}
where $\tilde{\delta}$ depends on the first moments of the channel.
\end{enumerate}

\noindent Notice that the displacement~(\ref{Displ_correct}) is equivalent
to set $\beta =\beta ^{\prime }-\tilde{\gamma}-\tilde{\delta}$ into the
Wigner function~(\ref{Cond_Bob_Wig2bis}) and consider the new variable $%
\beta ^{\prime }$. But, in order to simplify the notation, we can put $\beta
^{\prime }\equiv \beta $ and, therefore, the displacement~(\ref
{Displ_correct}) is equivalent to the formal substitution $\beta =\beta -%
\tilde{\gamma}-\tilde{\delta}$ into Eq.~(\ref{Cond_Bob_Wig2bis}). According
to the previous four-step protocol, the final state at Bob's station is
given by
\begin{equation}
W(\beta -\tilde{\gamma}-\tilde{\delta}|\tilde{\gamma})=\mathcal{P}(\tilde{%
\gamma})^{-1}\int d^{2}\gamma ~W^{in}(\gamma )W^{ch}(\gamma ^{\ast }-\tilde{%
\gamma}^{\ast },\beta -\tilde{\gamma}-\tilde{\delta})~.
\label{teleported_BOB}
\end{equation}
We may ask how much, on average, it is similar to the input state. In order
to get teleportation fidelity $F$, we compute the mean state teleported to
Bob by averaging over all possible results $\tilde{\gamma}$%
\begin{equation}
W^{out}(\beta ,\tilde{\delta})=\int d^{2}\tilde{\gamma}~\mathcal{P}(\tilde{%
\gamma})W(\beta -\tilde{\gamma}-\tilde{\delta}|\tilde{\gamma})~.
\label{Wout}
\end{equation}
Inserting Eq.~(\ref{teleported_BOB}) into Eq.~(\ref{Wout}), we get the
input-output relation
\begin{equation}
W^{out}(\beta ,\tilde{\delta})=\int d^{2}\gamma ~K(\beta -\gamma ,\tilde{%
\delta})W^{in}(\gamma )~,  \label{InOut}
\end{equation}
where the kernel
\begin{equation}
K(\beta -\gamma ,\tilde{\delta})=\int d^{2}\tilde{\gamma}~W^{ch}(\gamma
^{\ast }-\tilde{\gamma}^{\ast },\beta -\tilde{\gamma}-\tilde{\delta})~,
\label{Kernel}
\end{equation}
takes both the quantum channel $W^{ch}(\alpha ,\beta )$ and Bob's additional
displacement $\tilde{\delta}$ into account. If we now consider a \emph{pure}
state $\left| \psi \right\rangle _{in}$ as input, the fidelity (\ref
{Fid_pure}) takes the form\
\begin{equation}
F=\pi \int d^{2}\alpha ~W^{in}(\alpha )W^{out}(\alpha ),
\label{fid_PSpicture}
\end{equation}
and, by using the input-output relation~(\ref{InOut}), we get the final
formula
\begin{equation}
F(\tilde{\delta})=\pi \int d^{2}\alpha ~d^{2}\gamma ~K(\alpha -\gamma ,%
\tilde{\delta})[W^{in}(\alpha )]^{2},  \label{fid_delta}
\end{equation}
which expresses the fidelity as function of the input state, the channel and
Bob's additional displacement.

Let us consider the particular case of Gaussian states, i.e., a pure
Gaussian state as input and a two-mode Gaussian state as channel. Such
states are characterized by the two pairs $\mathbf{V}_{in},d_{in}$ and $%
\mathbf{V}_{ch},d_{ch}$, respectively. From Eq.~(\ref{fid_delta}) one can
prove that fidelity does not depend on the input displacement $d_{in}$ and,
therefore, it will depend on the input and the channel via $\mathbf{V}_{in},$
$\mathbf{V}_{ch}$ and $d_{ch}$, only. If we write $\mathbf{V}_{ch}$ in the
blockform of Eq.~(\ref{V_blocks_2modi}) and $d_{ch}\equiv 2\left(
d_{1},d_{2},d_{3},d_{4}\right) ^{T}$, from Eq.~(\ref{fid_delta}) we get
\begin{equation}
F(\tilde{\delta})=\frac{1}{\sqrt{\det \mathbf{\Gamma }}}\exp [-Q(\tilde{%
\delta})]~,  \label{Fidgen}
\end{equation}
where
\begin{equation}
\mathbf{\Gamma }\equiv 2\mathbf{V}_{in}+\mathbf{ZAZ}+\mathbf{B}-\mathbf{ZC}-%
\mathbf{C}^{T}\mathbf{Z}^{T}~,  \label{termGM}
\end{equation}
and
\begin{align}
Q(\tilde{\delta})& \equiv h(\tilde{\delta})^{T}\mathbf{\Gamma }^{-1}h(\tilde{%
\delta}),  \label{termQ} \\
h(\tilde{\delta})& \equiv \left( -\tilde{\delta}^{R}+d_{1}-d_{3,}-\tilde{%
\delta}^{I}-d_{2}-d_{4}\right) ^{T}.  \label{termh}
\end{align}
From Eq.~(\ref{termGM}) one can verify that the matrix $\mathbf{\Gamma }$ is
equal to
\begin{equation}
\mathbf{\Gamma }=\left(
\begin{array}{cc}
2\mathbf{V}_{in}^{11}+\langle \Delta \hat{X}_{-}^{2}\rangle & 2\mathbf{V}%
_{in}^{12}-\langle \Delta \hat{X}_{-}\Delta \hat{P}_{+}\rangle \\
2\mathbf{V}_{in}^{21}-\langle \Delta \hat{X}_{-}\Delta \hat{P}_{+}\rangle & 2%
\mathbf{V}_{in}^{22}+\langle \Delta \hat{P}_{+}^{2}\rangle
\end{array}
\right) ~,  \label{Gamma_Explicit}
\end{equation}
where $\hat{X}_{-}\equiv \hat{x}_{a}-\hat{x}_{b}$ and $\hat{P}_{+}\equiv
\hat{p}_{a}+\hat{p}_{b}$ are the EPR operators. Thus, one can easily prove
that $\mathbf{\Gamma }>0$ and $\det \mathbf{\Gamma }\geq 1$, so that $Q(%
\tilde{\delta})\geq 0$ in Eq.~(\ref{termQ}), and $F(\tilde{\delta})\leq 1$
in Eq.~(\ref{Fidgen}). Matrix $\mathbf{\Gamma }$ depends on CMs $\mathbf{V}%
_{in},\mathbf{V}_{ch}$ via Eq.~(\ref{termGM}), while the positive term $Q$
is linked to vector $h$ via Eq.~(\ref{termQ}), which, in turn, contains both
the shift of the channel and Bob's additional displacement $\tilde{\delta}$
(see Eq.~(\ref{termh})) Without such a displacement (i.e., $\tilde{\delta}=0$%
), teleported state acquires a nonzero shift from the channel and the
corresponding fidelity $F(0)$ will depend upon such a shift via a decreasing
exponential. In general, Bob can eliminate this shift by choosing an
additional displacement $\tilde{\delta}$ which perfectly cancels the effects
of $d_{ch}$ according to Eq.~(\ref{termh}), i.e.,
\begin{equation}
\tilde{\delta}^{R}=d_{1}-d_{3}\text{~\ \ \ and\ \ \ ~}\tilde{\delta}%
^{I}=-d_{2}-d_{4}~.  \label{termdelt2}
\end{equation}
In such a case, fidelity of Eq.~(\ref{Fidgen}) becomes
\begin{equation}
F=1/\sqrt{\det \mathbf{\Gamma }}~,  \label{Fidgen_CM}
\end{equation}
i.e., it becomes \emph{independent} from the displacement of the channel and
it takes the \emph{maximum} value, which is determined only by CMs $\mathbf{V%
}_{in}$ and $\mathbf{V}_{ch}$ via the matrix $\mathbf{\Gamma }$. Thus, for a
channel and an input state which are Gaussian, the ``right'' displacement at
Bob's station is given by Eq.~(\ref{Displ_correct}) with $\tilde{\delta}$
given in Eq.~(\ref{termdelt2}), and the corresponding expression for the
fidelity depends on the CMs of the channel and the input via the very simple
Eq.~(\ref{Fidgen_CM}).

\subsubsection{Fidelity, EPR\ correlations and entanglement}

Suppose now that our quantum channel is a Gaussian two-mode state with CM in
the normal form~(\ref{normal_FORM}), i.e., $\mathbf{V}_{ch}=\mathbf{V}^{%
\text{I}}(a,b,c,c^{\prime })$ and the input state is a coherent state, i.e.,
$\mathbf{V}_{in}=\mathbf{I}/2$. Computing the matrix $\mathbf{\Gamma }$ as
in Eq.~(\ref{termGM}) and inserting the result into Eq.~(\ref{Fidgen_CM}),
the optimal fidelity turns out to be
\begin{equation}
F^{(coh)}=\left[ \left( 1+\langle \Delta \hat{X}_{-}^{2}\rangle \right)
\left( 1+\langle \Delta \hat{P}_{+}^{2}\rangle \right) \right] ^{-1/2}~,
\label{FIDstform}
\end{equation}
where $\langle \Delta \hat{X}_{-}^{2}\rangle =a+b-2c$ and $\langle \Delta
\hat{P}_{+}^{2}\rangle =a+b+2c^{\prime }$. In particular, if $c^{\prime }=c$
then $\langle \Delta \hat{X}_{-}^{2}\rangle \geq 1$ and $\langle \Delta \hat{%
P}_{+}^{2}\rangle \geq 1$, so that $F^{(coh)}\leq 1/2=F_{class}^{(coh)}$,
i.e., teleportation is not quantum. On the contrary, if $c^{\prime }=-c$,
i.e., $\mathbf{V}_{ch}=\mathbf{V}^{\text{I}}(a,b,c,-c)$, then we have $%
\langle \Delta \hat{X}_{-}^{2}\rangle =\langle \Delta \hat{P}_{+}^{2}\rangle
\equiv \aleph $ as in Eq.~(\ref{EPR_measure}), where quantity $\aleph $\
represents a measure of the EPR\ correlations of the relevant quadratures
of\ the channel. From Eq.~(\ref{FIDstform}) it follows that
\begin{equation}
F^{(coh)}=\left[ 1+\aleph \right] ^{-1}~,  \label{FID_EPRchannel}
\end{equation}
which shows that (for the considered class of channels) the existence of EPR
correlations is equivalent to quantum teleportation, i.e.,
\begin{equation}
\aleph <1\Leftrightarrow F^{(coh)}>\frac{1}{2}=F_{class}^{(coh)}~.
\label{EPReqTEL}
\end{equation}
A Gaussian state described by a CM $\mathbf{V}^{\text{I}}(a,b,c,-c)$ with $%
\aleph <1$ represents a Gaussian EPR channel\textit{\ }(see Sec.~\ref
{EPR_CHANNEL}) and generalizes the two-parameter EPR channel studied in
Ref.~ \cite{ClassicF}. For this kind of channel, one can simply prove
\begin{equation}
F^{(coh)}>1/2~\Longrightarrow \text{~\emph{entanglement}}~\text{\emph{.}}
\label{CS_Ent_Fid}
\end{equation}
But implication~(\ref{CS_Ent_Fid}) holds in general for every quantum
channel \cite{F1su2}. In fact, a teleportation protocol fully based on
classical strategies is exactly a teleportation process which does not use
any quantum entanglement between Alice and Bob, and since it implies a
fidelity $F^{(coh)}\leq 1/2$ for coherent states' teleportation, the
condition $F^{(coh)}>1/2$ is sufficient for the existence of bipartite
quantum entanglement.

An important one-parameter Gaussian EPR channel is the TMSV state, with the
squeezing-dependent CM $\mathbf{V}(r)$ specified in Eq.~(\ref{CM_TB}). In
such a case $\aleph =\exp (-2r)$, and therefore
\begin{equation}
F^{(coh)}(r)=e^{2r}(1+e^{2r})^{-1}~,
\end{equation}
which gives $1/2$ for $r=0$ while $F^{(coh)}(r)\longrightarrow 1$ for $%
r\longrightarrow +\infty $. In this limit of infinite squeezing, the TMSV
state becomes an ideal EPR\ pair and allows to implement a perfect quantum
teleportation. In real experiments one can access only finite squeezing ($%
0<r<+\infty $) and realize a quantum teleportation with fidelity $%
1/2<F^{(coh)}(r)<1$ like, e.g., in Ref.~\cite{Furusawa}, where the
experimental value $F^{(coh)}=0.58\pm 0.02$\ has been reported.

In general, it is an open problem to connect the entanglement of the quantum
channel, shared by two parties, with the fidelity of teleportation of
coherent states. In particular, one would like to show a monotonic relation
between the fidelity and an entanglement monotone, like, e.g., the
log-negativity or the minimum PT symplectic eigenvalue. Recently, Ref.~\cite
{AdessoTele} showed that this is possible in the case of a thermal TMSV
state, which can be described by the CM
\begin{equation}
\mathbf{V}(r,n_{a},n_{b})=\frac{1}{2}\left(
\begin{array}{cccc}
n_{a}e^{2r}+n_{b}e^{-2r} & 0 & n_{a}e^{2r}-n_{b}e^{-2r} & 0 \\
0 & n_{a}e^{-2r}+n_{b}e^{2r} & 0 & n_{a}e^{-2r}-n_{b}e^{2r} \\
n_{a}e^{2r}-n_{b}e^{-2r} & 0 & n_{a}e^{2r}+n_{b}e^{-2r} & 0 \\
0 & n_{a}e^{-2r}-n_{b}e^{2r} & 0 & n_{a}e^{-2r}+n_{b}e^{2r}
\end{array}
\right) ,  \label{Themal_TMSV}
\end{equation}
where $n_{a}$ and $n_{b}$ are thermal noises which enters the system via
modes $a$ and $b$. Alice and Bob can try to fully exploit the resources of
this channel by performing suitable LOCCs just before teleportation. For
instance, they can apply local squeezing transformations, described in phase
space by $\mathbf{V\longrightarrow V}^{\prime }\equiv \mathbf{SVS}^{T}$
where $\mathbf{S}=\mathrm{diag}(e^{\kappa },e^{-\kappa },e^{\kappa
},e^{-\kappa })$. These local symplectic transformations do not alter the
entanglement, i.e., the minimum PT symplectic eigenvalue $\tilde{\nu}_{-}$
remains unchanged from $\mathbf{V}$ to $\mathbf{V}^{\prime }$, while the
fidelity $F^{(coh)}$, computed via Eqs.~(\ref{termGM}) and~(\ref{Fidgen_CM}%
), will depend on the parameter $\kappa $. One can then prove \cite
{AdessoTele} that $F^{(coh)}$ takes a maximum for $\kappa =[\log
(n_{a}/n_{b})]/4$, for which we have
\begin{equation}
F_{\max }^{(coh)}=1/(1+\tilde{\nu}_{-}).  \label{fid_PTeigenv}
\end{equation}
Thus, if Alice and Bob share a thermal TMSV state, the maximal teleportation
fidelity for coherent states, achieved by optimizing over local squeezing
transformations, is connected to the entanglement of the underlying quantum
channel by the simple Eq.~(\ref{fid_PTeigenv}), which shows that $F_{\max
}^{(coh)}$ is an entanglement monotone for this particular class of states.

\section{Continuous variable quantum teleportation networks\label{INTRO_NET}%
\label{CV_networks}}

\begin{figure}[tbph]
\vspace{-0cm}
\par
\begin{center}
\includegraphics[width=0.7\textwidth]{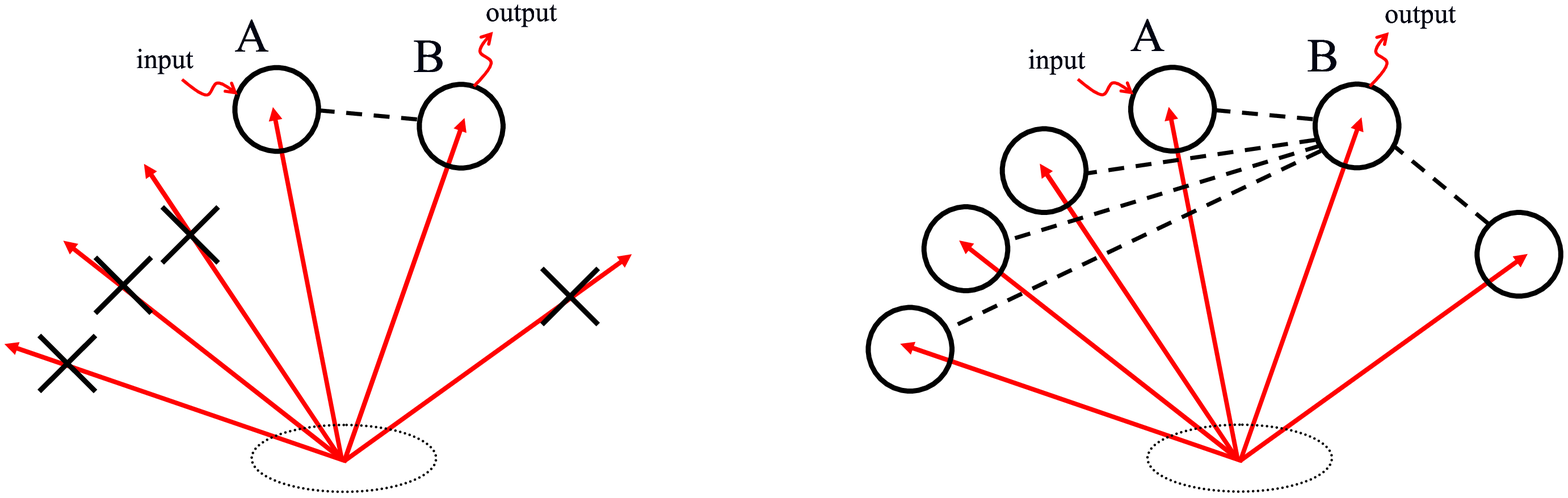}
\end{center}
\par
\vspace{-0.4cm}
\caption{\textbf{Quantum teleportation network}. In the \emph{non-assisted }%
protocol (left), a pair of nodes of the network (Alice and Bob) ignores all
the other nodes and\ performs a standard teleportation protocol. In the
\emph{assisted} protocol (right), all the other nodes take part to the
process by means of LOCCs. They perform suitable local measurements and then
classically communicate their outcomes to Bob. Bob, in turn, uses this
additional information to perform a modified transformation.}
\label{NetN}
\end{figure}

In the previous sections, we have recognized quantum entanglement as an
important resource for quantum information purposes. In particular,
bipartite entanglement represents the fundamental requirement that a shared
quantum channel should have in order to enable a truly quantum
teleportation. The possibility to generate multipartite entanglement, i.e.,
entanglement shared by more than two parties \cite{cinjap,Lance, FuruNat},
has recently opened up the possibility to construct \emph{quantum
teleportation networks}. Here, a set of \emph{nodes} have in common a
quantum channel which enables to perform quantum teleportation between an
arbitrary pair in the net. In a first strategy, the teleportation between
these two arbitrary nodes can be implemented in the standard way (see Fig.~%
\ref{NetN}), which corresponds to ignore all the other nodes and exploit the
residual bipartite entanglement together with classical communications. This
strategy is a direct extension of the standard teleportation protocol from
two to more stations and we call it \emph{non-assisted} protocol. But a more
clever strategy is based upon a cooperative behavior, where all the other
nodes assist the teleportation between the chosen pair (Alice and Bob) by
means of LOCCs. In fact, if the external nodes perform suitable local
measurements and then classically communicate their outcomes to Bob, the
latter can use this additional classical information to improve the process
via modified conditional displacements. We call this strategy \emph{assisted}
protocol.

In the CV domain, the idea of a CV quantum teleportation network was firstly
developed in Ref.~\cite{BraNetwork}, where an easy way to produce CV
multipartite entanglement was shown. The recipe of Ref.~\cite{BraNetwork} to
create multipartite entanglement relies on a suitable distribution of
squeezing. In the case of a single mode, a squeezed state $\left| \alpha
,\varepsilon \right\rangle $\ is determined by two parameters: the complex
amplitude $\alpha $ and the complex squeeze factor $\varepsilon \equiv r\exp
(2i\varphi )$, which is, in turn, composed by a real \emph{squeeze factor} $%
r $ and a \emph{squeeze phase} $\varphi $~$(0\leq \varphi \leq \pi )$. Such
a state can be generated by squeezing the vacuum and then by displacing it,
i.e., $\left| \alpha ,\varepsilon \right\rangle =\hat{D}(\alpha )\hat{S}%
(\varepsilon )\left| 0\right\rangle $ where
\begin{equation}
\hat{S}(\varepsilon )=\hat{S}(r,\varphi )=\exp \tfrac{1}{2}(\varepsilon
^{\ast }\hat{a}^{2}-\varepsilon \hat{a}^{\dagger 2})
\end{equation}
is the one-mode squeezing operator, and
\begin{equation}
\hat{D}(\alpha )=\exp (\alpha \hat{a}^{\dagger }-\alpha ^{\ast }\hat{a})
\end{equation}
is the displacement operator (which expresses the Weyl operator in terms of
complex amplitude). When two squeezed states are taken as input to a beam
splitter, the squeezing mixes between the two modes and creates EPR
correlations among their quadratures. For this reason, the state of the two
output modes is entangled and such a procedure can be suitably extended from
two to $N\geq 2$ modes.

Let us consider a set of $N$ modes which will be then distributed to $N$
different \emph{nodes} of a teleportation network. Before distribution, one
can create a multipartite entangled state by applying local squeezing
transformations to these $N$ modes, followed by a collective \emph{N-splitter%
} transformation \cite{Nsplitter}, which is a suitable sequence of beam
splitter transformations. Denote by $\left| 0\right\rangle _{k}$ the initial
vacuum state of the arbitrary mode $k$ ($k=1,...,N$), and define the \emph{%
N-splitter} transformation as
\begin{equation}
\hat{N}_{1...N}\equiv \hat{B}_{N-1,N}(\pi /4)\hat{B}_{N-2,N-1}(\cos ^{-1}1/%
\sqrt{3})\times ...\times \hat{B}_{1,2}(\cos ^{-1}1/\sqrt{N})~,
\label{N_Splitter}
\end{equation}
where
\begin{equation}
\hat{B}_{i,j}(\theta ):\hat{a}_{i}\longrightarrow \hat{a}_{i}\cos \theta +%
\hat{a}_{j}\sin \theta ,~\hat{a}_{j}\longrightarrow \hat{a}_{i}\sin \theta -%
\hat{a}_{j}\cos \theta  \label{BS_transformation}
\end{equation}
corresponds to a lossless beam splitter operation on the pair of modes $i$
and $j$. Then, we can generate the following multipartite state
\begin{equation}
\left| \Psi (r)\right\rangle =\hat{N}_{1...N}\left[ \hat{S}_{1}(r,\varphi
=\pi /2)\left| 0\right\rangle _{1}\otimes \left( \bigotimes_{k=2}^{N}\hat{S}%
_{k}(r,\varphi =0)\left| 0\right\rangle _{k}\right) \right] ~,~r>0,
\label{Multi_channel}
\end{equation}
which is constructed by squeezing mode $1$ in momentum and the other modes $%
2,...,N$ in position, and then by applying the \emph{N-splitter}~(\ref
{N_Splitter}) to all of them. The final state~(\ref{Multi_channel}) is a
Gaussian state with Wigner function \cite{cvbook}
\begin{equation}
W(\zeta )=\left( \frac{2}{\pi }\right) ^{N}\exp \left\{ -\tfrac{\xi ^{-1}}{2N%
}\left[ 2\left( \sum_{k=1}^{N}x_{k}\right) ^{2}+\sum_{k,l=1}^{N}\left(
p_{k}-p_{l}\right) ^{2}\right] -\tfrac{\xi }{2N}\left[ 2\left(
\sum_{k=1}^{N}p_{k}\right) ^{2}+\sum_{k,l=1}^{N}\left( x_{k}-x_{l}\right)
^{2}\right] \right\} ~,  \label{Multi_channel_Wig}
\end{equation}
where $\xi \equiv \exp (2r)>1$ and $\zeta ^{T}=(x_{1},p_{1},...,x_{N},p_{N})$
as usual. The multipartite state~(\ref{Multi_channel_Wig}) is totally
symmetric under interchange of nodes, so that every pair of them, among the $%
N(N-1)/2$ different choices in the network, has the same resources for
quantum teleportation. One can prove \cite{cvbook} that the state~(\ref
{Multi_channel_Wig}) implies bipartite entanglement for every pair of the
net, when all the other nodes are traced out. For this reason, it is a good
quantum channel which, potentially, enables to perform quantum teleportation
between an arbitrary pair of the net. It is straightforward to prove that,
for $N=2$, the Wigner function~(\ref{Multi_channel_Wig}) is equal to the
Wigner function~(\ref{Wig_TB}) of the TMSV state, where the input squeezings
(parameters $(r,\pi /2)$ and $(r,0)$ in Eq.~(\ref{Multi_channel}) for $N=2$)
are directly transformed in two-mode squeezing/EPR correlations (parameter $%
r $ in Eq.~(\ref{CM_TB})). Practically, the state of Eq.~(\ref{Multi_channel}%
), corresponding to the Wigner function of Eq.~(\ref{Multi_channel_Wig}),
represents the \emph{multi-mode} version of the TMSV state, which we call
\emph{N-mode squeezed vacuum} (NMSV)\ state. Furthermore, if we take the
limit of infinite squeezing $\xi \longrightarrow +\infty $ in Eq.~(\ref
{Multi_channel_Wig}), such Wigner function becomes infinitely peaked
at\thinspace $p_{1}+...+p_{N}=0$ and $x_{k}-x_{l}=0$ (for $k,l=1,2,...,N$),
i.e., it becomes the \emph{multi-mode} version of the EPR state of Eq.~(\ref
{limite_EPR})
\begin{equation}
W(\zeta )\longrightarrow W_{EPR}(\zeta )\propto \delta
(p_{1}+...+p_{N})\prod_{k,l=1}^{N}\delta (x_{k}-x_{l})~.
\label{limit_MD_EPR}
\end{equation}
This ideal \emph{N-mode} EPR state can be rigorously defined as the \emph{%
zero-value} eigenstate of the total momentum $\sum_{k}\hat{p}_{k}$ and
relative positions $\hat{x}_{k}-\hat{x}_{l}$. In the Heisenberg picture, it
corresponds to the following conditions for the quadratures
\begin{equation}
\sum_{k}\hat{p}_{k}=0,~\hat{x}_{k}-\hat{x}_{l}=0~(\forall k,l)~.
\label{EPR_Nmode_Heis}
\end{equation}
The usage of an ideal \emph{N-mode} EPR state, as shared quantum channel
within a \emph{N-node} CV quantum teleportation network (\emph{N-mode}
teleportation network, in the following), represents the ideal asymptotic
situation corresponding to infinite squeezing, where one can exploit the
perfect EPR correlations and perform perfect quantum teleportations within
the network. But, actually, one can access only finite squeezing and,
therefore, one can only exploit a multipartite channel as the NMSV state of
Eq.~(\ref{Multi_channel}) that has a finite amount of squeezing $r$. In this
real situation the performances of the teleportation processes are clearly
lower.

In the following we will first present the case of an ideal CV quantum
teleportation within the network, adopting the Heisenberg picture for
simplicity. In particular, we will discuss two different protocols: the
\emph{non-assisted} protocol, which is connected to the notion of \emph{%
telecloning}, and the \emph{assisted} protocol, which exploits homodyne
detection in the ideal case. Then, we will deal with a more realistic
situation, where the quantum channel has not perfect EPR\ correlations. In
such a case, teleportation fidelity is not equal to one, and it is open
issue to determine what are the LOCCs that optimize the fidelity of the
assisted protocol. This is the central problem analyzed in the subsequent
sections. From now on, for simplicity, we consider only the case $N=3$.

\subsection{Ideal three-mode teleportation network\label{IDEAL_3MTN}}

In the ideal three-mode teleportation network the quantum channel is given
by an ideal \emph{3-mode} EPR state. Teleportation can be performed by two
arbitrarily chosen nodes of the net, the sender (Alice) and the receiver
(Bob), and the protocol can be \emph{assisted} or \emph{non-assisted} by the
third node (Charlie). In the first case, a suitable homodyne detection of
the third node allows to achieve a perfect teleportation ($F=1$), while this
is not possible in the latter case, due to the unexploited quantum resources
owned by the third node. In this case, the third node may attempt to use his
part of quantum channel to perform an additional teleportation from the
sender to him (\emph{telecloning}). In the Heisenberg picture we have the
following protocol(s):

\begin{enumerate}
\item  \emph{Initial condition}. Three distant parties, identified with
Alice, Bob and Charlie, possess three different modes $a,b$ and $c$,
respectively, which have been prepared in the ideal \emph{3-mode} EPR\ state
\begin{equation}
\hat{x}_{a}-\hat{x}_{b}=\hat{x}_{a}-\hat{x}_{c}=\hat{x}_{b}-\hat{x}_{c}=0~,~~%
\hat{p}_{a}+\hat{p}_{b}+\hat{p}_{c}=0~.  \label{EPR_corr3}
\end{equation}
These three parties compose an ideal three-mode teleportation network since
they can exploit their shared perfect EPR correlations~(\ref{EPR_corr3}) to
implement an instance of perfect teleportation within the network. Suppose
that one of the three, say Alice, wants to teleport to one of the other two,
say Bob, an unknown state of an input mode $in$ with quadratures $\hat{x}%
_{in}$ and $\hat{p}_{in}$.

\item  \emph{Bell measurement}. Alice realizes the CV version of the Bell
measurement on her modes by the usual two steps:

\begin{enumerate}
\item  \emph{Beam splitter mixing}. Alice mixes the $in$ mode with her $a$
mode (part of the channel) via a balanced beam splitter
\begin{equation}
\hat{x}_{\pm }=(\hat{x}_{a}\pm \hat{x}_{in})/\sqrt{2}~,~\hat{p}_{\pm }=(\hat{%
p}_{a}\pm \hat{p}_{in})/\sqrt{2}~,  \label{XP_BSplitter_3modes}
\end{equation}
where $\hat{x}_{\pm }$ and $\hat{p}_{\pm }$ are the quadratures of the
output modes $\pm $ .

\item  \emph{Homodyne detection}. Alice detects quadratures $\hat{x}_{-}$
and $\hat{p}_{+}$. Denoting with $(x_{-},p_{+})$ the outcomes, her
measurement causes in Eq.~(\ref{XP_BSplitter_3modes}) the usual collapse
\begin{equation}
\hat{x}_{a}=\hat{x}_{in}+\sqrt{2}x_{-}~,~\hat{p}_{a}=-\hat{p}_{in}+\sqrt{2}%
p_{+}~.  \label{XP_collapse_3modes}
\end{equation}
Due to the EPR property~(\ref{EPR_corr3}), Bob and Charlie's quadratures are
instantaneously projected according to (\ref{XP_collapse_3modes}), i.e., we
have
\begin{align}
\hat{x}_{b}& =\hat{x}_{in}+\sqrt{2}x_{-}~,~\hat{p}_{b}=\hat{p}_{in}-\sqrt{2}%
p_{+}-\hat{p}_{c}~,  \label{Bob_coll_3modes} \\
\hat{x}_{c}& =\hat{x}_{in}+\sqrt{2}x_{-}~,~\hat{p}_{c}=\hat{p}_{in}-\sqrt{2}%
p_{+}-\hat{p}_{b}~.  \label{Charlie_coll_3modes}
\end{align}
As we can see from Eq.~(\ref{Bob_coll_3modes}), Bob's momentum $\hat{p}_{b}$%
\ contains now an additional momentum operator $\hat{p}_{c}$ with respect to
Eq.~(\ref{Bob_collapse}). The presence of such an operator hints that
Charlie can take an active part into the process and, at this point of the
protocol, we may distinguish between \emph{assisted} protocol and \emph{%
non-assisted} protocol.
\end{enumerate}
\end{enumerate}

\begin{itemize}
\item  \underline{Non-assisted protocol}. Charlie does not intervene in the
process. Consequently, Alice and Bob ignores Charlie, and accomplish the
teleportation according to the standard protocol shown in Sec.~\ref
{Ideal_CV_QT}:
\end{itemize}

\begin{enumerate}
\item[3]  \emph{Classical communication}. Alice communicates her measurement
result $(x_{-},p_{+})$ to Bob through the classical channel.

\item[4]  \emph{Conditional displacement}. Bob uses this classical
information $(x_{-},p_{+})$ to perform the usual conditional displacement
\begin{equation}
\begin{array}{l}
\hat{x}_{b}\longrightarrow \hat{x}_{b}^{\prime }\equiv \hat{x}_{b}-\sqrt{2}%
x_{-}=\hat{x}_{in}~, \\
\hat{p}_{b}\longrightarrow \hat{p}_{b}^{\prime }\equiv \hat{p}_{b}+\sqrt{2}%
p_{+}=\hat{p}_{in}-\hat{p}_{c}~.
\end{array}
\label{Bob_displ_3modes}
\end{equation}
Contrary to Eq.~(\ref{Bob_displ}), in Eq.~(\ref{Bob_displ_3modes}) Bob does
not obtain exactly the input momentum $\hat{p}_{in}$. In the Schroedinger
picture, it means that the output state is not exactly the input one, but it
has a certain fidelity $\tilde{F}<1$ because of the quantum resources owned
by Charlie. What is the value of this fidelity? In order to answer to this
question, assume that Alice has broadcasted her measurement result $%
(x_{-},p_{+})$. In such a case, also Charlie receives this classical
information and can attempt to teleport Alice's input state to himself. From
Eq.~(\ref{Charlie_coll_3modes}), we see that the best local operation he can
do is the same of Bob, i.e., the conditional displacement
\begin{equation}
\begin{array}{l}
\hat{x}_{c}\longrightarrow \hat{x}_{c}^{\prime }\equiv \hat{x}_{c}-\sqrt{2}%
x_{-}=\hat{x}_{in}~, \\
\hat{p}_{c}\longrightarrow \hat{p}_{c}^{\prime }\equiv \hat{p}_{c}+\sqrt{2}%
p_{+}=\hat{p}_{in}-\hat{p}_{b}~.
\end{array}
\label{Charlie_displ_3modes}
\end{equation}
Thus, independently and via the same shared quantum channel, both Bob and
Charlie perform an \emph{imperfect} teleportation from Alice. At the end of
the whole process, the input state at Alice's station is disappeared, and
two imperfect copies appear at Bob and Charlie's stations. On average (i.e.,
repeating the process many times with the same input) such copies are
identical, but they cannot be identical to the original input state because
of the \emph{no-cloning theorem} \cite{NoCloning}. For this reason, both the
Alice-Bob and the Alice-Charlie teleportations have the same fidelity $%
\tilde{F}$\ which must satisfy the bound $\tilde{F}\leq F^{1\longrightarrow
2}$, where $F^{1\longrightarrow 2}$ is the optimal $1\longrightarrow 2$\
cloning fidelity for the states considered in input (for instance $%
F^{1\longrightarrow 2}=2/3$ for coherent states \cite{NoCloCohe}). The whole
protocol, where Bob and Charlie do not assist each other and perform two
independent teleportations from Alice, represents the standard example of $%
\emph{telecloning}$ protocol \cite{CVteleCLO}. In general, for telecloning
we mean a process where an unknown input state is attempted to be teleported
to $N$ different recipients so that $N$ (imperfect) clones are created at
the remote locations.
\end{enumerate}

\begin{itemize}
\item  \underline{(Ideal) assisted protocol}. Charlie gets involved in the
process with the aim of helping Bob. According to Eq.~(\ref{Bob_coll_3modes}%
), the best thing he can do is to measure momentum $\hat{p}_{c}$ and
communicate the result $p_{c}$. In fact, this additional information,
besides the one communicated by Alice, allows Bob to perform a \emph{%
modified }conditional displacement on his mode $b$\ which establishes,
again, $\hat{x}_{in},\hat{p}_{in}$ at the output. In detail the assisted
protocol goes as follows:
\end{itemize}

\begin{enumerate}
\item[3']  \emph{Assisting measurement}. Charlie performs a homodyne
detection on his mode $c$ which realizes the measurement of the momentum $%
\hat{p}_{c}$. Denoting by $p_{c}$ the outcome, from Eq.~(\ref
{Bob_coll_3modes}) it follows the collapse
\begin{equation}
\hat{x}_{b}=\hat{x}_{in}+\sqrt{2}x_{-}~,~\hat{p}_{b}=\hat{p}_{in}-\sqrt{2}%
p_{+}-p_{c}~.  \label{collapse_Charlie}
\end{equation}

\item[4']  \emph{Classical communication}. Alice and Charlie communicate
their measurement results, $(x_{-},p_{+})$ and $p_{c}$, to Bob through a
classical channel.

\item[5']  \emph{Conditional displacement}. Bob uses the total classical
information $(x_{-},p_{+},p_{c})$ to perform a modified conditional
displacement on his own mode $b$%
\begin{equation}
\begin{array}{l}
\hat{x}_{b}\longrightarrow \hat{x}_{b}^{\prime }\equiv \hat{x}_{b}-\sqrt{2}%
x_{-}=\hat{x}_{in}~, \\
\hat{p}_{b}\longrightarrow \hat{p}_{b}^{\prime }\equiv \hat{p}_{b}+\sqrt{2}%
p_{+}+p_{c}=\hat{p}_{in}~,
\end{array}
\label{Bob_displ_3m_Ass}
\end{equation}
which allows to complete the teleportation process. In fact, according to
Eq.~(\ref{Bob_displ_3m_Ass}), mode $b$ is finally described by the conjugate
quadratures of the input. In the Schroedinger picture this is equivalent to
have teleported the state from Alice to Bob with fidelity $F=1$.
\end{enumerate}

\subsection{Real three-mode teleportation networks\label{Real Nets}}

In the previous section, we have considered an ideal three-mode
teleportation network, as generated by an ideal three-mode EPR state.
Actually, in any real experimental setup, one can produce only finite
squeezing and, therefore, imperfect EPR correlations. The corresponding
\emph{real} quantum channels imply teleportation processes with lower
performances, i.e., with $F<1$, even if it can be very close to one.
Remarkable examples of real quantum channels are the 3MSV state, i.e., the
NMSV state~(\ref{Multi_channel}) with $N=3$, and the following \emph{cheap}
three-mode state
\begin{equation}
\left| \Phi (r)\right\rangle =\hat{N}_{123}\left[ \hat{S}_{1}(r,\varphi =\pi
/2)\left| 0\right\rangle _{1}\otimes \left| 0\right\rangle _{2}\otimes
\left| 0\right\rangle _{3}\right] ~,~r>0~,  \label{Brau_econ}
\end{equation}
which is generated by squeezing \emph{only one} mode before the \emph{%
3-splitter}. The latter channel is sufficient to allow quantum teleportation
between every pair of nodes in the net since, for every $r>0$, it gives $%
F^{(coh)}>1/2$ for teleportation of coherent states, if the assisted
protocol\ is adopted \cite{BraNetwork}. In addition to these quantum
channels realized using squeezers and beam splitters, there are other
examples of three-mode quantum channels, as the ones generated by
inter-linked bilinear interactions \cite{Paris} or by radiation pressure
within opto-mechanical systems \cite{PRL,PRAtele}. Also in these cases\
quantum teleportation within the networks have been proved to be feasible
and an assisted strategy, consisting in a LOCC by Charlie, can be proved to
improve the process. It must be stressed that such a LOCC must not be
necessarily based upon a homodyne detection at Charlie's site, as it happens
in the ideal case (and also in the case of the cheap state~(\ref{Brau_econ})
and the 3MSV state). In general, when we consider general kinds of channels,
there is the non trivial open question to determine what is the best LOCC
that Charlie must perform, in order to optimize teleportation fidelity of
the assisted protocol. Actually, such question is very general, and we will
specialize it to the case of LOCCs given by local \emph{measurements} at
Charlie's site. Then we will consider only the\emph{\ Gaussian} case, i.e.,
shared quantum channels and input states which are Gaussian states. In
detail, the open problem which we want to solve is

\begin{enumerate}
\item[\textbf{Problem}]  \emph{In a three-mode network where the quantum
channel is given by an arbitrary three-mode Gaussian state, what is
Charlie's local measurement (and classical communication) which optimizes
the fidelity of teleportation of pure Gaussian states between Alice and Bob?}
\end{enumerate}

\noindent In the following section we attempt to solve this problem of the
assisted protocol. From a mathematical point of view, we can face the
problem exploiting connections with the derivation of Sec.~\ref{REAL_CV_QT}.
Let us denote by $\rho $ the three-mode Gaussian state shared by Alice, Bob
and Charlie. The non-assisted protocol is equivalent to trace Charlie and
apply the above derivation to the two-mode reduced state $\rho ^{tr}\equiv
\mathrm{Tr}_{c}(\rho )$, i.e., assuming $\rho ^{tr}$ as\ quantum channel $%
\rho _{ab}$ between Alice and Bob. The assisted protocol, where Charlie
performs a local measurement and communicates the result $n$ to Bob, can be
treated by adopting that derivation with $\rho _{ab}=\rho ^{(n)}$, where $%
\rho ^{(n)}$ is the two-mode conditional state created by Charlie's
measurement.

\subsection{Optimal local measurement\label{OPTIMAL_POVM}}

Consider a three-mode network where Alice, Bob and Charlie share a quantum
channel given by a three-mode Gaussian state (\emph{three-mode Gaussian
network} from now on). Such quantum channel is used to perform quantum
teleportation of a pure Gaussian state between two of the parties (Alice and
Bob), while the third party (Charlie) can condition the process by means of
LOCCs. Here, we show the best measurement that Charlie can perform on his
own mode that preserves the Gaussian character of the three-mode state and
optimizes the teleportation fidelity between Alice and Bob \cite{POVMpra}.
In other words, we find the optimal Gaussian measurement at Charlie's site
which maximizes the teleportation fidelity of the assisted protocol.
Motivated by related results, we conjecture that this optimal Gaussian
measurement is also the best among \emph{all} possible measurements, trying
to give a general answer to the problem of Sec.~\ref{Real Nets}. In Sec.~\ref
{Ass_Not_Ass_PROTO} we present the scenario and describe the teleportation
protocols, assisted and not assisted by measurements at Charlie's site. In
Sec.~\ref{DICHOTOMIC_POVM} we discuss the case when Charlie performs a
dichotomic measurement with a Gaussian and a non-Gaussian outcome, while in
Sec.~\ref{LOCAL_G_POVM} we consider the case of a local Gaussian
measurement. For simplicity, we have omitted the proofs of the theorems
which are reported in Ref.~\cite{POVMpra}.

\subsubsection{Assisted and non-assisted teleportation protocols\label%
{Ass_Not_Ass_PROTO}}

\begin{figure}[ptbh]
\vspace{-1.0cm}
\par
\begin{center}
\includegraphics[width=1\textwidth]{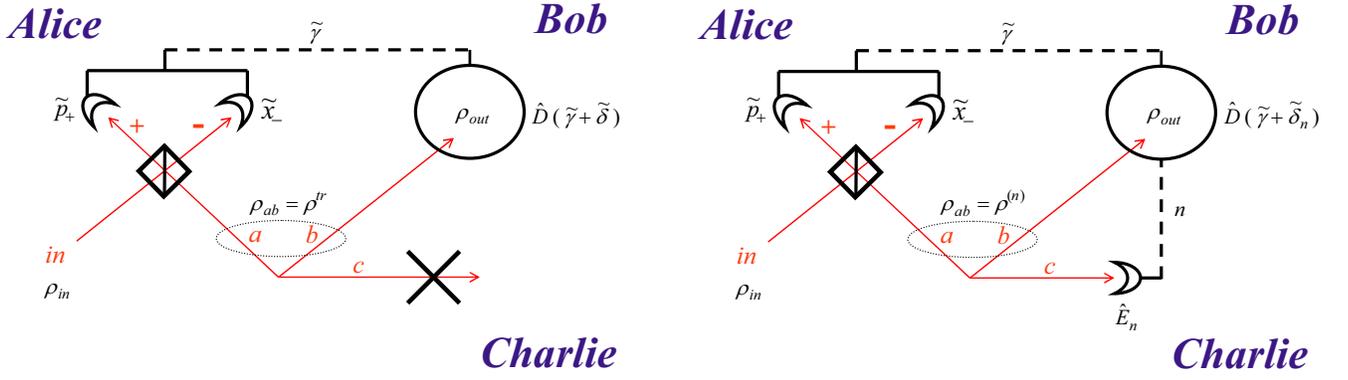}
\end{center}
\par
\vspace{-0.8cm} \caption{An arbitrary 3-mode Gaussian state is
shared by Alice, Bob and
Charlie. Alice is supplied with an unknown pure Gaussian state $\protect\rho%
_{in}$ which she wants to teleport to Bob. In a first non-assisted strategy
(left picture), Charlie is traced out and Alice and Bob implement a standard
continuous variable teleportation protocol. In an alternative assisted
strategy (right picture), Bob is helped by Charlie who detects his mode and
classically communicates the result $n$ to Bob, who uses also this
information for his local operation. Here we consider, for Charlie's
measurement, first a local dichotomic measurement and then a local Gaussian
measurement.}
\label{AssNotAss}
\end{figure}

Let us consider a three-mode Gaussian network where Alice has to teleport an
unknown pure Gaussian state to Bob. In a first strategy, Charlie is
neglected, and Alice and Bob implement the standard CV\ teleportation
protocol of Sec.~\ref{REAL_CV_QT} (\emph{non-assisted} protocol). In an
alternative strategy, Charlie performs a local measurement and classically
communicates the result to Bob, who exploits this additional information in
the teleportation process (\emph{assisted} protocol).\ Here, we show the two
protocols in detail (see also Fig.~\ref{AssNotAss}). For both protocols, the
various steps are basically the same as presented in Sec.~\ref{IDEAL_3MTN},
but, here, we prefer to show them giving the precedence to Charlie's action
(nothing in the first case, a LOCC in the second case). The inversion of the
order between Alice and Charlie's physical actions is possible because of
the local character of their measurements, and it allows to study the
process applying directly the results of Sec.~\ref{REAL_CV_QT}.

\begin{enumerate}
\item  \emph{Initial condition}. Alice, Bob and Charlie possess three
different modes, characterized by annihilation operators $\hat{a},\hat{b}$
and $\hat{c}$ respectively, which are described by a three-mode Gaussian
state $\rho $, having displacement $d^{T}\equiv
(d_{a}^{T},d_{b}^{T},d_{c}^{T})\in {\mathbb{R}}^{6}$ and CM
\begin{equation}
\mathbf{V}\equiv \left(
\begin{array}{ccc}
\mathbf{A} & \mathbf{F} & \mathbf{E} \\
\mathbf{F}^{T} & \mathbf{B} & \mathbf{D} \\
\mathbf{E}^{T} & \mathbf{D}^{T} & \mathbf{C}
\end{array}
\right) \mathbf{~},  \label{CMtot}
\end{equation}
where the blocks $\mathbf{A},\mathbf{B},...,\mathbf{F}$ are $2\times 2$ real
matrices. Alice has to teleport to Bob a single-mode pure Gaussian state $%
\rho _{in}$ with CM $\mathbf{V}_{in}$ and displacement $d_{in}\in {\mathbb{R}%
}^{2}$ completely unknown to her. Since a single-mode pure Gaussian state is
a single-mode squeezed state, we can put $\rho _{in}=\left| \alpha
_{in},\varepsilon _{in}\right\rangle \left\langle \alpha _{in},\varepsilon
_{in}\right| $.
\end{enumerate}

\begin{itemize}
\item  \underline{Non-assisted protocol}. The most straightforward strategy
is to ignore Charlie (see Fig.~\ref{AssNotAss}, left side) and use the
reduced two-mode state $\rho ^{tr}\equiv \mathrm{Tr}_{c}(\rho )$ to
implement a standard CV teleportation protocol. It is easy to prove that
such a state is a Gaussian state with CM
\begin{equation}
\mathbf{V}^{tr}=\left(
\begin{array}{cc}
\mathbf{A} & \mathbf{F} \\
\mathbf{F}^{T} & \mathbf{B}
\end{array}
\right) ~,  \label{CMtr}
\end{equation}
and displacement $(d^{tr})^{T}=(d_{a}^{T},d_{b}^{T})\in {\mathbb{R}}^{4}$.
Thus, we can apply the results of Sec.~\ref{REAL_CV_QT}, relative to
Gaussian states, by taking $\rho _{ab}=\rho ^{tr}$.
\end{itemize}

\begin{enumerate}
\item[2.]  \emph{Bell measurement}. Alice mixes her part of the channel
(mode $a$)\ with the input (mode $in$) through a balanced beam-splitter and
makes a homodyne detection of the output modes, i.e., she measures the
quadratures $\hat{x}_{-}\equiv 2^{-1/2}(\hat{x}_{a}-\hat{x}_{in})$ and $\hat{%
p}_{+}\equiv 2^{-1/2}(\hat{p}_{a}+\hat{p}_{in})$. As result she obtains $(%
\tilde{x}_{-},\tilde{p}_{+})$ which can be compacted in the complex number $%
\tilde{\gamma}\equiv -\tilde{x}_{-}+i\tilde{p}_{+}$.

\item[3.]  \emph{Classical communication}. Alice communicates the result $%
\tilde{\gamma}$ to Bob through a classical channel.

\item[4.]  \emph{Conditional displacement}. Bob uses this classical
information $\tilde{\gamma}$ to perform the conditional displacement $\hat{D}%
(\tilde{\gamma}+\tilde{\delta})$ on his mode $b$, where $\tilde{\gamma}$
compensates Alice's measurement and $\tilde{\delta}$ the shift $d^{tr}$ of
the channel according to Eq.~(\ref{termdelt2}) (with $d_{ch}=d^{tr}$).
Thanks to this displacement, Bob achieves a shift-independent fidelity\
which takes its maximum value. In particular, it depends on the CMs $\mathbf{%
V}_{in}$ and $\mathbf{V}^{tr}$\ of the input and reduced channel according
to Eqs.~(\ref{termGM}) and~(\ref{Fidgen_CM}), i.e., we have the \emph{%
non-assisted} fidelity
\begin{equation}
F^{tr}=(\det \mathbf{\Gamma }^{tr})^{-1/2}~,  \label{Fidel_tr}
\end{equation}
where
\begin{equation}
\mathbf{\Gamma }^{tr}\equiv 2\mathbf{V}_{in}+\mathbf{ZAZ}+\mathbf{B}-\mathbf{%
ZF}-\mathbf{F}^{T}\mathbf{Z}^{T}~.  \label{gammatr}
\end{equation}
\end{enumerate}

\begin{itemize}
\item  \underline{Assisted protocol}. An alternative strategy for Alice and
Bob is to ask for the help of Charlie, who can perform a suitable
measurement on his own mode $c$ and classically communicate the result to
Bob (see Fig.~\ref{AssNotAss}, right side). In this modified protocol, Bob
performs his displacement only after he has received the information about
the measurement outcomes from \emph{both} Alice and Charlie.
\end{itemize}

\begin{enumerate}
\item[2'.]  \emph{Assisting LOCC}. Charlie performs a local measurement on
his mode $c$, obtaining an outcome $n$ with probability $P_{n}$. Then, he
communicates the result $n$ to Bob through a classical channel. At this
point, the resulting quantum channel between Alice and Bob is described by a
reduced two-mode state $\rho ^{(n)}$ conditioned to the outcome $n$, i.e., $%
\rho _{ab}=\rho ^{(n)}$. Such a state can be Gaussian or not, it depends on
the outcome and on the kind of measurement at Charlie's site. Since the
quantum channel is conditional, all the teleportation process is
conditional, and consequently we define a fidelity $F^{(n)}$ conditioned to
the outcome $n$. This corresponds to the fidelity of the process if the
outcome $n$ was always selected by Charlie's measurement (remember that the
fidelity is a mean quantity, averaged over the measurements of the
protocol). But the measurement is probabilistic, and therefore the effective
fidelity of the process will be given, at the end, by $F=%
\sum_{n}P_{n}F^{(n)} $, which we call \emph{assisted} fidelity. This
fidelity is no more conditioned to the particular outcome of the assisting
measurement, but it is still a conditional quantity, since it is conditioned
to the kind of assisting measurement chosen by Charlie.

\item[3'.]  \emph{Bell measurement} and \emph{classical communication}.
After that Charlie has created the conditional quantum channel $\rho ^{(n)}$%
\ by means of his LOCC, Alice continues the protocol exactly as previous
points 2 and 3, i.e., she performs the Bell measurement and classically
communicates the result $\tilde{\gamma}$\ to Bob.

\item[4'.]  \emph{Conditional displacement}. Bob uses both the classical
datas $n$ (from Charlie)\ and $\tilde{\gamma}$ (from Alice), to
perform a conditional displacement $\hat{D}(\tilde{\gamma}+\tilde{\delta}%
_{n})$ on his mode $b$, where $\tilde{\gamma}$ compensates Alice's
measurement and $\tilde{\delta}_{n}$ balances the shift $d^{(n)}$ of the
conditional reduced channel $\rho ^{(n)}$. In general, there exists an
optimal displacement $\tilde{\delta}_{n}$ which optimizes the conditional
fidelity $F^{(n)}$ according to Eq.~(\ref{fid_delta}). Thanks to such a
displacement, Bob can maximize the conditional fidelity $F^{(n)}$\ for every
outcome $n$, and therefore he can optimize the effective fidelity $%
F=\sum_{n}P_{n}F^{(n)}$ of the assisted protocol. In the particular
instances where $\rho ^{(n)}$ is a Gaussian state (CM $\mathbf{V}^{(n)}$ and
displacement $d^{(n)}$), Bob can compute such a displacement directly from
Eq.~(\ref{termdelt2}) (with $d_{ch}=d^{(n)}$). The corresponding \emph{%
Gaussian} conditional fidelity takes the form (see Eq.~(\ref{Fidgen_CM}))
\begin{equation}
F^{(n)}=(\det \mathbf{\Gamma }^{(n)})^{-1/2}~,
\end{equation}
where $\mathbf{\Gamma }^{(n)}$ depends on the CMs $\mathbf{V}_{in}$ and $%
\mathbf{V}_{ch}=\mathbf{V}^{(n)}$ according to Eq.~(\ref{termGM}).
\end{enumerate}

In the following we consider two general kinds of quantum measurements at
Charlie's site: a local dichotomic measurement, with a Gaussian outcome and
a non-Gaussian one, and a local Gaussian measurement,\ defined as a local
measurement preserving the Gaussian character of the shared state for every
outcome. Our aim is to compare the assisted fidelity $F$ and the
non-assisted fidelity $F^{tr}$ for both kinds of measurement. We anticipate
that for the dichotomic measurement one does not have an improvement ($F\leq
F^{tr}$), but the results achieved for the conditional fidelities $F^{(n)}$\
are very useful and they can be directly extended to the case of the
Gaussian measurement, where one can optimize $F$ and surely state that $%
F\geq F^{tr}$.

\subsubsection{Dichotomic measurement\label{DICHOTOMIC_POVM}}

We first consider the case of a dichotomic measurement with operators $\hat{E%
}_{0},\hat{E}_{1}\equiv (\hat{I}-\hat{E}_{0}^{2})^{1/2}$ where $\hat{E}_{0}$
is an arbitrary Gaussian state with CM $\mathbf{V}_{0}$ and displacement $%
d_{0}$. This implies that for the outcome $n=0$ the conditional reduced
state $\rho ^{(0)}\equiv P_{0}^{-1}\mathrm{Tr}_{c}(\hat{E}_{0}\rho \hat{E}%
_{0}^{\dagger })$ is still Gaussian. After some algebra \cite{POVMpra}, one
can prove that $\rho ^{(0)}$ is Gaussian having CM
\begin{equation}
\mathbf{V}^{(0)}=\mathbf{V}^{tr}-\left(
\begin{array}{cc}
\mathbf{EME}^{T} & \mathbf{EMD}^{T} \\
\mathbf{DME}^{T} & \mathbf{DMD}^{T}
\end{array}
\right) ~,  \label{CM0}
\end{equation}
and displacement
\begin{equation}
d^{(0)}=d^{tr}+\left(
\begin{array}{c}
\mathbf{EM}(d_{0}-d_{c}) \\
\mathbf{DM}(d_{0}-d_{c})
\end{array}
\right) ~.  \label{drift_red}
\end{equation}
In the previous Eqs.~(\ref{CM0}) and~(\ref{drift_red}), $\mathbf{M}$ is the
following real, symmetric and strictly positive matrix
\begin{equation}
\mathbf{M}\equiv g^{-1}\mathbf{J}\left[ 2(\det \mathbf{V}_{0}+1/4)\mathbf{V}%
_{0}+4(\det \mathbf{V}_{0})\mathbf{C}\right] \mathbf{J}^{T}~,  \label{M}
\end{equation}
and
\begin{equation}
g\equiv 4\det \mathbf{V}_{0}\det \mathbf{C}+2(\det \mathbf{V}_{0}+1/4)%
\mathrm{Tr}(\mathbf{V}_{0}\mathbf{JCJ}^{T})+(\det \mathbf{V}%
_{0}+1/4)^{2}>1/2~.  \label{fattoreG}
\end{equation}
Since the conditional reduced state $\rho ^{(0)}$\ is Gaussian, the
corresponding conditional fidelity of the assisted protocol is given by
\begin{equation}
F^{(0)}=(\det \mathbf{\Gamma }^{(0)})^{-1/2}~,  \label{fidel_0}
\end{equation}
where the strictly positive matrix $\mathbf{\Gamma }^{(0)}$ depends on CMs $%
\mathbf{V}_{in}$ and $\mathbf{V}_{ch}=\mathbf{V}^{(0)}$\ according to Eq.~(%
\ref{termGM}). Using Eq.~(\ref{CM0}) we obtain
\begin{equation}
\mathbf{\Gamma }^{(0)}=\mathbf{\Gamma }^{tr}-\mathbf{\Sigma }^{T}\mathbf{%
M\Sigma }~,  \label{gamma0}
\end{equation}
where
\begin{equation}
\mathbf{\Sigma }\equiv \mathbf{E}^{T}\mathbf{Z}-\mathbf{D}^{T}~,
\label{sigma}
\end{equation}
and $\mathbf{\Gamma }^{tr}$\ is given in Eq.~(\ref{gammatr}). Since the
conditional reduced state $\rho ^{(0)}$\ is Gaussian, the conditional
reduced state $\rho ^{(1)}$ corresponding to the other outcome $n=1$ is
necessarily non-Gaussian since $\rho ^{(1)}=P_{1}^{-1}[\rho ^{tr}-P_{0}\rho
^{(0)}]$. In correspondence of the \emph{non-Gaussian outcome} $n=1$, Bob
performs its local displacement using a suitable additional shift $\delta
^{(1)}$ which aims to optimize the conditional fidelity $F^{(1)}$ according
to the formula of Eq.~(\ref{fid_delta}). Totally, the assisted fidelity $F$,
i.e., the effective fidelity of the teleportation protocol assisted by the
dichotomic measurement, is given by $F=P_{0}F^{(0)}+P_{1}F^{(1)}$. One can
prove \cite{POVMpra} the following theorem, which specifies the relationship
among the non-assisted fidelity $F^{tr}$\ and the conditional/assisted
fidelities in the case of the dichotomic measurement

\begin{description}
\item[\textbf{Theorem (monotonicity).}]  The conditional fidelities $F^{(n)}$%
\ corresponding to the outcomes $n=0,1$\ and the assisted fidelity $F$\
satisfy the inequality
\begin{equation}
F^{(1)}\leq F\leq F^{tr}\leq F^{(0)}~.  \label{result1}
\end{equation}
\end{description}

\noindent Eq.~(\ref{result1}) shows that the dichotomic measurement leads,
on average, to an assisted fidelity $F$ which does not outperform $F^{tr}$,
proving that the present dichotomic scheme does not seem to bring advantages
in a real teleportation network. However, the situation is very interesting
from the point of view of the conditional teleportation fidelities $F^{(n)}$%
. In fact Eq.~(\ref{result1}) shows that teleportation fidelity always
increases if Charlie performs a measurement and the corresponding
conditional state is still Gaussian ($F^{(0)}\geq F^{tr}$), while it always
decreases with respect to the trace case for the outcome corresponding to
the non-Gaussian conditional state ($F^{(1)}\leq F^{tr}$), even if in this
latter case the conditional bipartite state is more pure than that without
measurement \cite{nota}. This result suggests which is the right kind of
measurement to be considered at Charlie's site (local Gaussian measurement)
and it will be the starting point of the next Section~\ref{LOCAL_G_POVM}.

Actually the dichotomic scheme can bring advantages as a
\emph{probabilistic} scheme, where Bob asks Alice to perform the
Bell measurement and the classical communication only if Charlie's
measurement has given the Gaussian outcome. In such a case the
assisted fidelity $F$ is just the conditional one $F^{(0)}$, but
the protocol has a success probability equal to $P_{0}$. One also
verify that, in some instances, the Gaussian outcome $n=0$ can
give $F^{(0)}>1/2$ for the teleportation of coherent states when
$\rho ^{tr}$ is not entangled (and therefore $F^{tr}\leq 1/2$). In
other words, Charlie can conditionally generate remote bipartite
entanglement between Alice and Bob if the dichotomic measurement
selects the Gaussian outcome. This process corresponds to an
\emph{entanglement localization} as already investigated in
Refs.~\cite{LOCALI,LOCALI2} for qubits and CV systems. All these
considerations make clear why it is profitable to optimize the
\emph{Gaussian} conditional fidelity $F^{(0)}$ upon the
measurement parameters, and exactly such optimization work
concerns the remainder of this section.

Thus, we restrict to the Gaussian outcome ($n=0$), and look for the optimal
Gaussian state $\hat{E}_{0}$ which maximizes the fidelity $F^{(0)}$. As a
first result, one can prove \cite{POVMpra} the following

\begin{description}
\item[\textbf{Theorem (purity).}]  \emph{\label{THEO_PURITY}}For every
Gaussian state $\hat{E}_{0}$, there exists a pure Gaussian state $\hat{E}%
_{0,p}$\ such that $F^{(0,p)}\geq F^{(0)}$.
\end{description}

\noindent According to the latter result, the optimal Gaussian measurement
operator $\hat{E}_{0}$ is actually a projection onto a pure Gaussian state,
and therefore it has to be searched within the set of squeezed states $%
\left| \alpha ,\varepsilon \right\rangle =\hat{D}(\alpha )\hat{S}%
(\varepsilon )\left| 0\right\rangle $. Assuming that Charlie knows the CMs $%
\mathbf{V}$ and $\mathbf{V}_{in}$, he can optimize the fidelity $F^{(0)}$
with respect to the CM $\mathbf{V}_{0}$ of the Gaussian operator $\hat{E}%
_{0} $ (the displacement $d_{0}$\ can be arbitrary). According to the
previous purity theorem, he can accomplish this task by optimizing\ directly
$F^{(0,p)}$ with respect to the CM\ of the arbitrary squeezed state $\left|
\alpha ,\varepsilon \right\rangle $, which is given by
\begin{equation}
\mathbf{V}_{0}(\xi ,\varphi )=\frac{1}{2}\left(
\begin{array}{cc}
\xi \sin ^{2}\varphi +\xi ^{-1}\cos ^{2}\varphi & (\xi -\xi ^{-1})\sin
\varphi \cos \varphi \\
(\xi -\xi ^{-1})\sin \varphi \cos \varphi & \xi \cos ^{2}\varphi +\xi
^{-1}\sin ^{2}\varphi
\end{array}
\right) ~,  \label{V0}
\end{equation}
where $\xi \equiv \exp (2r)$. Thanks to the two-parameter form of the CM~(%
\ref{V0}), the optimization has been reduced to an optimization over $\xi $
and $\varphi $. However, finding a global maximum point $(\bar{\xi},\bar{%
\varphi})$ is difficult in general, and we must split the problem in two
steps: we first maximize the fidelity $F^{(0,p)}\equiv F(\xi ,\varphi )$
with respect to $\xi $\ for an arbitrary but fixed $\varphi $, and then we
maximize the result $F(\bar{\xi}(\varphi ),\varphi )$ with respect to $%
\varphi $. Mathematically speaking the function $F(\xi ,\varphi )$ is
bounded and continuous in the domain $]0,+\infty \lbrack \times \lbrack
0,\pi ]$ and we implicitly have to consider its continuous extension in $%
[0,+\infty ]\times \lbrack 0,\pi ]$ in order to surely have the existence of
global extremal points. At the end of the computation, Charlie determines an
optimal couple $(\bar{\xi},\bar{\varphi})\in \lbrack 0,+\infty ]\times
\lbrack 0,\pi ]$ describing a \emph{finitely}, or eventually \emph{infinitely%
}, squeezed state which optimizes the Gaussian conditional fidelity $F^{(0)}$%
. In both cases we can denote the optimal Gaussian measurement operator
compactly\ as $\hat{E}_{0}^{opt}=\left| \bar{\xi},\bar{\varphi},\alpha
\right\rangle \left\langle \bar{\xi},\bar{\varphi},\alpha \right| $ with $%
\alpha $ arbitrary. In particular, for $\bar{\xi}=+\infty $ it becomes the
eigenstate $\hat{E}_{0}^{opt}=\left| x(\bar{\varphi})\right\rangle
\left\langle x(\bar{\varphi})\right| $, and for $\bar{\xi}=0$ it becomes the
eigenstate $\hat{E}_{0}^{opt}=\left| x(\bar{\varphi}+\pi /2)\right\rangle
\left\langle x(\bar{\varphi}+\pi /2)\right| $, where $\hat{x}(\varphi
)\equiv 2^{-1/2}(\hat{c}\,e^{-i\varphi }+\hat{c}^{\dagger }\,e^{i\varphi })$
and the corresponding eigenvalue $x$ is arbitrary. Note that, here, we
include the infinitely squeezed states within the set of Gaussian states.
This can be justified for two reasons: first, they are a sort of \emph{%
asymptotic} Gaussian states, since limits of Gaussian state; then, they
generate a conditional reduced state $\rho ^{(0)}$ which is still Gaussian
\cite{IntroGauss}.

The first step of maximization is solved by the following

\begin{description}
\item[\textbf{Theorem (optimization).}]  For every squeezing phase $\varphi
\in \lbrack 0,\pi ]$, Charlie can select a squeezing factor $\bar{\xi}%
(\varphi )$ such that $F(\bar{\xi}(\varphi ),\varphi )\geq F(\xi ,\varphi )$
for every $\xi $ (phase-dependent global maximum point). The point $\bar{\xi}%
(\varphi )$ can be derived analytically from the CMs $\mathbf{V}$ and $%
\mathbf{V}_{in}$, according to the following four-step procedure:
\end{description}

\begin{enumerate}
\item  Construct the matrices $\mathbf{\Gamma }^{tr}$ of Eq.~(\ref{gammatr}%
), $\mathbf{\Sigma }$ of Eq.~(\ref{sigma}) and $\mathbf{U}\equiv \mathbf{%
\Sigma J\Gamma }^{tr}\mathbf{J}^{T}\mathbf{\Sigma }^{T}$.

\item  {Define the 2D vectors }
\begin{equation}
u\equiv \left(
\begin{array}{c}
u_{x} \\
u_{y}
\end{array}
\right) =\left(
\begin{array}{c}
\det \mathbf{C}+1/4 \\
(\det \mathbf{\Sigma })^{2}-\mathrm{Tr}(\mathbf{JCJ}^{T}\mathbf{U})
\end{array}
\right) ,  \label{vector_u}
\end{equation}
{and }
\begin{equation}
{k(\varphi )\equiv \left(
\begin{array}{c}
k_{x}(\varphi ) \\
k_{y}(\varphi )
\end{array}
\right) =\left(
\begin{array}{c}
{\vartheta }^{T}\mathbf{U}{\vartheta } \\
{\vartheta }^{T}\mathbf{C}{\vartheta }
\end{array}
\right) ,}  \label{vector_k}
\end{equation}
{where $\vartheta ^{T}\equiv (\sin \varphi ,\cos \varphi )$.}

\item  {Consider the real numbers $\gamma (\varphi )\equiv u^{T}k(\varphi )$
and $\omega (\varphi )\equiv \lbrack k_{x}(\varphi )k_{y}(\varphi -\pi
/2)-k_{y}(\varphi )k_{x}(\varphi -\pi /2)]/2$.}

\item  {Denote by $p(\varphi )$ the $\varphi $-dependent logic proposition $%
\gamma (\varphi )<0~$}$\wedge ~${$\gamma (\varphi -\pi /2)<0$ .}

Then
\begin{align}
p(\varphi )& =1\Longleftrightarrow \bar{\xi}(\varphi )=\frac{\omega (\varphi
)-\sqrt{\omega (\varphi )^{2}+\gamma (\varphi -\pi /2)\gamma (\varphi )}}{%
\gamma (\varphi -\pi /2)}\equiv \xi _{-}(\varphi )~,  \label{sol1} \\
p(\varphi )& =0\Longleftrightarrow \bar{\xi}(\varphi )=0\text{\quad }\vee
\quad \bar{\xi}(\varphi )=+\infty ~.  \label{sol2}
\end{align}
\end{enumerate}

\noindent The second step concerns the maximization over the squeezing phase
$\varphi $. Using the two vectors of Eqs.~(\ref{vector_u}) and~(\ref
{vector_k}),\ the fidelity can be written as
\begin{equation}
F(\xi ,\varphi )=\left[ \det \mathbf{\Gamma }^{tr}-\frac{-u_{y}+(\xi
/2)k_{x}(\varphi -\pi /2)+(\xi ^{-1}/2)k_{x}(\varphi )}{u_{x}+(\xi
/2)k_{y}(\varphi -\pi /2)+(\xi ^{-1}/2)k_{y}(\varphi )}\right] ^{-1/2}.
\label{Fvector}
\end{equation}
We then consider the piecewise continuous function of $\varphi $, $\xi =\bar{%
\xi}(\varphi )$ defined according to (\ref{sol1}), (\ref{sol2})\ and the
corresponding phase-dependent teleportation fidelity $\bar{F}(\varphi
)\equiv F(\bar{\xi}(\varphi ),\varphi )$ which is continuous on $[0,\pi ]$.
From Eq.~(\ref{V0}) one has
\begin{equation}
\mathbf{V}_{0}(\xi ,\varphi )=\mathbf{V}_{0}(\xi ,\varphi +\pi )=\mathbf{V}%
_{0}(\xi ^{-1},\varphi +\pi /2)~,  \label{periodicity}
\end{equation}
and therefore $F(0,\varphi )=F(+\infty ,\varphi +\pi /2)$ and $F(0,\varphi
+\pi /2)=F(+\infty ,\varphi )$. This implies that finding the maximum point $%
\bar{\varphi}$ of $\bar{F}(\varphi )$ is equivalent to find the maximum
point of the piecewise continuous function
\begin{equation}
\tilde{F}(\varphi )=\left\{
\begin{array}{c}
F(\xi _{-}(\varphi ),\varphi )\qquad \quad \qquad \qquad \qquad \qquad
\qquad \mathrm{if}\quad p(\varphi )=1\quad , \\
F(0,\varphi )=[\det \mathbf{\Gamma }^{tr}-k_{x}(\varphi )/k_{y}(\varphi
)]^{-1/2}\qquad \mathrm{if}\quad p(\varphi )=0\quad .
\end{array}
\right.  \label{Ftilde}
\end{equation}
Now, one has three cases: (i) $\bar{\varphi}$ is a stationary point of $%
F(0,\varphi )$; (ii) $\bar{\varphi}$ is a stationary point of $F(\xi
_{-}(\varphi ),\varphi )$; (iii) $\bar{\varphi}$ is one of the border points
dividing the intervals where $p(\varphi )=1$ from those where $p(\varphi )=0$%
. We report a simple analytical expression of the final global maximum point
only in the first case, while in the other two cases the expressions are
extremely involved. In case (i), defining the $2\times 2$\ matrix $%
\mbox{
       \boldmath{
           \small{\!\!\!\!\!$\tau $}}}\equiv \mathbf{UJCJ}^{T}$, the
stationary points $\varphi _{\pm }$ of $F(0,\varphi )$ are given by the
relation
\begin{equation}
\cos 2\varphi _{\pm }=\frac{%
\mbox{
       \boldmath{
           \small{\!\!\!\!\!$\tau $}}}_{12}^{2}-%
\mbox{
       \boldmath{
           \small{\!\!\!\!\!$\tau $}}}_{21}^{2}\pm (%
\mbox{
       \boldmath{
           \small{\!\!\!\!\!$\tau $}}}_{11}-%
\mbox{
       \boldmath{
           \small{\!\!\!\!\!$\tau $}}}_{22})\sqrt{(%
\mbox{
       \boldmath{
           \small{\!\!\!\!\!$\tau $}}}_{11}-%
\mbox{
       \boldmath{
           \small{\!\!\!\!\!$\tau $}}}_{22})^{2}+4%
\mbox{
       \boldmath{
           \small{\!\!\!\!\!$\tau $}}}_{12}%
\mbox{
       \boldmath{
           \small{\!\!\!\!\!$\tau $}}}_{21}}}{(%
\mbox{
       \boldmath{
           \small{\!\!\!\!\!$\tau $}}}_{11}-%
\mbox{
       \boldmath{
           \small{\!\!\!\!\!$\tau $}}}_{22})^{2}+(%
\mbox{
       \boldmath{
           \small{\!\!\!\!\!$\tau $}}}_{12}+%
\mbox{
       \boldmath{
           \small{\!\!\!\!\!$\tau $}}}_{21})^{2}}~.  \label{FiMax}
\end{equation}
In many cases of practical interest (for instance when coherent states or $%
\varphi =0$ squeezed states are teleported through a CM
$\mathbf{V}$ with diagonal blocks, as for example in Refs.~\cite
{BraNetwork,CVteleCLO,Paris,PRL,PRAtele,game}), the above
procedure allows to find the maximum point
$(\bar{\xi}(\bar{\varphi}),\bar{\varphi})$
and the corresponding optimal conditional fidelity $F_{max}^{(0)}=F(\bar{\xi}%
(\bar{\varphi}),\bar{\varphi})$ quite quickly.\ In some easy cases, when
matrices $\mathbf{U}$ and $\mathbf{C}$ are proportional to the identity, the
above optimization becomes $\varphi $-independent, and therefore the maximum
point is given by $\bar{\xi}=1$ if $\gamma <0$ or by $\bar{\xi}=0$ if $%
\gamma \geq 0$. In the first case the optimal Gaussian measurement\ operator
is a coherent state, i.e., $\hat{E}_{0}^{opt}=\left| \alpha \right\rangle
\left\langle \alpha \right| $ (with $\alpha $ arbitrary), while in the
second case it is an infinitely squeezed state, i.e., $\hat{E}%
_{0}^{opt}=\left| x(\varphi )\right\rangle \left\langle x(\varphi )\right| $
(with phase $\varphi $ and eigenvalue $x(\varphi )$ arbitrary).

\subsubsection{Local Gaussian measurement\label{LOCAL_G_POVM}}

The above optimization results refer to the conditional scheme where Charlie
performs a dichotomic measurement and the Gaussian outcome $n=0$ is
selected. However, these results can be extended to a different scheme where
all the measurement outcomes are Gaussian and, therefore, all the
conditional fidelities can potentially outperform $F^{tr}$ according to the
monotonicity theorem. More in detail, such theorem suggests to consider a
measurement which\ creates a conditional bipartite state which is Gaussian
for every outcome. We surely achieve this condition if Charlie performs a
local Gaussian measurement, i.e., a local measurement $\{\hat{E}(n)\}$
transforming a Gaussian multipartite state into another Gaussian state for
every outcome $n$. As before, here we consider the term \emph{Gaussian} in
an extended sense, i.e., including asymptotic Gaussian states as the
infinitely squeezed states. Examples of local Gaussian measurement are
heterodyne or homodyne measurements on mode $c$, or they are obtained when
the $c$ mode is coupled to ancillary modes by a Gaussian unitary interaction
(i.e., a LUBO) and the ancillas are subject to heterodyne or homodyne
measurements.

Consider then the assisted protocol where Charlie performs a local Gaussian
measurement $\{\hat{E}(n)\}$ and communicates the outcome $n$. It is
possible to prove a result analogous to purity theorem for the dichotomic
case:

\begin{description}
\item[\textbf{Theorem.}]  \label{THEO_L_GAUSS}For every local Gaussian
measurement $\{\hat{E}(n)\}$\ with fidelity $F$, there exists a ``pure''
local Gaussian measurement $\{\hat{E}_{\varepsilon }(\alpha )\equiv \left|
\alpha ,\varepsilon \right\rangle \left\langle \alpha ,\varepsilon \right| /%
\sqrt{\pi },\alpha \in \mathbb{C}\}$\ with a suitable $\varepsilon $, such
that the corresponding fidelity $F(\varepsilon )\geq F$.
\end{description}

\noindent Trivially the previous theorem assures the existence of a local
Gaussian measurement of the \emph{pure} form $\{\hat{E}_{\varepsilon
}(\alpha )\equiv \left| \alpha ,\varepsilon \right\rangle \left\langle
\alpha ,\varepsilon \right| /\sqrt{\pi },\alpha \in \mathbb{C}\}$ which
leads to an assisted fidelity $F(\varepsilon )\geq F^{tr}$. In fact\ it is
sufficient to consider $\{\hat{E}(n)\}=I$ and apply the theorem. More
importantly it implies that the optimal local Gaussian measurement must be
searched within the set of \emph{pure} measurements $\{\{\hat{E}%
_{\varepsilon }(\alpha )\},\varepsilon \in \mathbb{C}\}$, which is
equivalent to maximize with respect to the $2\times 2$ CM $\mathbf{V}%
_{0}(\varepsilon )=\mathbf{V}_{0}(\xi ,\varphi )$ of Eq.~(\ref{V0}). Thanks
to this result, the optimization procedure is exactly the one given for the
dichotomic measurement, i.e., it is given by the maximization over the
squeezing factor and the squeezing phase. Repeating such procedure, it is
possible to find an optimal pair of parameters $(\bar{\xi}(\bar{\varphi}),%
\bar{\varphi})$ which describes the optimal local Gaussian measurement $\{%
\hat{E}_{\bar{\varepsilon}}(\alpha )\}$ and provides the corresponding
optimal assisted fidelity $F(\bar{\xi},\bar{\varphi})$ via Eq.~(\ref{Fvector}%
). Notice that for finite squeezing ($\bar{\xi}\neq 0,+\infty $) the
measurement $\{\hat{E}_{\bar{\varepsilon}}(\alpha )\}$ can be realized by
first applying a unitary squeezing transformation $\hat{S}(\bar{\varepsilon}%
) $ to mode $c$\ and then making heterodyne detection. For infinite
squeezing ($\bar{\xi}=0,+\infty $) the measurement $\{\hat{E}_{\bar{%
\varepsilon}}(\alpha )\}$ is instead equivalent to a homodyne detection,
i.e., to $\left| x(\bar{\varphi})\right\rangle _{c}\left\langle x(\bar{%
\varphi})\right| $ for $\bar{\xi}=+\infty $ and to $\left| x(\bar{\varphi}%
+\pi /2)\right\rangle _{c}\left\langle x(\bar{\varphi}+\pi /2)\right| $ for $%
\bar{\xi}=0$, where $\hat{x}(\varphi )\equiv 2^{-1/2}(\hat{c}\,e^{-i\varphi
}+\hat{c}^{\dagger }\,e^{i\varphi })$ as usual. As for the dichotomic case,
we can use this optimized measurement to create conditional bipartite
entanglement between Alice and Bob, and now this can be done in a
deterministic way since all the outcomes are Gaussian.

\subsubsection{A conjecture \textit{vs} an open problem}

In the previous sections, we have studied how a two-party teleportation
process (between Alice and Bob) within a three-party shared quantum channel
can be conditioned by a local measurement and a classical communication of
the third party (Charlie). In particular, our analysis has been carried out
for a shared Gaussian channel, the teleportation of pure Gaussian states,
and two general kinds of local measurement at Charlie's site. We have first
shown the case of a dichotomic measurement and we have proved that the
non-Gaussian outcome always worsens the fidelity while the Gaussian outcome
always improves it, even allowing the conditional generation of
entanglement. Then we have shown how the dichotomic measurement can be
designed so that the Gaussian outcome optimizes the teleportation fidelity,
and we have extended such results directly to the case of a local Gaussian
measurement at Charlie's site. From the knowledge of the correlation
matrices (the one of the shared tripartite Gaussian state and the one of the
state to be teleported), Charlie can always determine and perform an optimal
local Gaussian measurement, given by a set of squeezed states with squeezing
factor $\bar{\xi}(\bar{\varphi})$ and squeezing phase $\bar{\varphi}$,
maximizing the fidelity of teleportation of pure Gaussian states between
Alice and Bob.

Let us come back to the problem of Sec.~\ref{Real Nets}. Our optimization
procedure allows us to give it a \emph{partial} answer, i.e., we can solve
such an open problem if we specialize it to the case of local \emph{Gaussian}
measurements

\begin{description}
\item[\textbf{Partial Solution}]  \emph{Let us consider a three-mode network
with quantum channel given by an arbitrary three-mode Gaussian state. From
the knowledge of the CMs (of the channel and the input), Charlie can always
determine and perform an optimal local \underline{Gaussian} measurement,
given by a set of squeezed states with squeezing factor $\bar{\xi}(\bar{%
\varphi})$ and squeezing phase $\bar{\varphi}$, maximizing the teleportation
fidelity of pure Gaussian states between Alice and Bob.}
\end{description}

\noindent It remains an open question to establish if this optimal Gaussian
measurement is also the best among \emph{all} the possible measurements at
Charlie's site. On the other hand, the monotonicity theorem clearly shows
that, in the dichotomic case, the Gaussian outcome yields always a better
result than the non-Gaussian one. This fact brings us to make the following
conjecture as attempt to address the open problem of Sec.~\ref{Real Nets}

\begin{description}
\item[\textbf{Conjecture}]  \emph{In a three-mode network with quantum
channel given by an arbitrary three-mode Gaussian state, the optimal local
Gaussian measurement $\{\hat{E}_{\bar{\varepsilon}}(\alpha )\}$, with
parameters $(\bar{\xi},\bar{\varphi})$ computed from the CMs of the channel
and the input as before, is Charlie's best local measurement in order to
optimize the teleportation fidelity of pure Gaussian states between Alice
and Bob.}
\end{description}

\noindent Such a conjecture is quite judicious if we meditate that, here, we
are considering the particular task of teleporting a single-mode pure \emph{%
Gaussian} state employing a tripartite \emph{Gaussian} state. Moreover, we
are implicitly considering among the local Gaussian measurements also \emph{%
asymptotic }Gaussian measurement operators, as in the case of the homodyne
detection.

\section{Conclusion\label{Conclusion}}

Quantum information theory with continuous variables provides an interesting
alternative to the traditional qubit-based approach, and it seems to be
particularly advantageous for quantum communications, as in the case of
quantum teleportation. Here, we have rapidly introduced some of the
mathematical tools to treat CV quantum systems and Gaussian states. Then, we
have extended the basic concepts of entanglement and teleportation to the
CV\ framework where we have reviewed some important connections between
fidelity, EPR\ correlations and entanglement measure. Such connections have
been shown for particular cases, since they are not known in general and
they are currently object of investigation.

As a generalization of CV quantum teleportation, we have approached the
study of CV quantum teleportation networks, where three or more users share
a multimode state and quantum teleportation between an arbitrary pair of
users can be assisted or non-assisted by all the other parties. Here,
assisted protocols have been proved to outperform non-assisted ones,
provided that a suitable set of LOCCs are performed by the assisting
parties. However, it is an open problem to determine what are the best LOCCs
which assist quantum teleportation. We have attempted to address this
problem in the case of a three-mode teleportation network, where the
performance of quantum teleportation between Alice and Bob conditioned to a
local measurement and classical communication by Charlie has been analyzed.
Such analysis refers to Gaussian states and could be eventually extended to
networks with more users.


\begin{thebibliography}{99}
\bibitem{relation}  In fact, in this case $E_{\mathcal{N}}(\rho _{AB})=\log
[(1+3x)/2]$ where $E_{\mathcal{N}}$ is the logarithmic negativity \cite
{VidalW}.
\bibitem{notaEPR}  The choice of such projectors depends on the EPR property
of Eq.~(\ref{CVIDEAL1}). Equivalently, one could have an EPR\ pair of the
type $\hat{x}_{A}+\hat{x}_{B}=\hat{p}_{A}-\hat{p}_{B}=0$. In such a case
Alice detects the other couple $\hat{x}_{+}$ and $\hat{p}_{-}$, and the
remainder of the protocol suffers of a change of sign.
\bibitem{nota}  Conditioning reduces entropy i.e $S(\rho ^{(n)})\leq S(\rho
^{tr})$ $\forall n$, where $S(\rho )=-Tr(\rho \log \rho )$ is the Von
Neumann entropy.
\end{thebibliography}
\end{document}